\documentclass[journal]{IEEEtran}%
\usepackage[utf8]{inputenc}
\usepackage{eurosym}
\usepackage{todonotes}
\usepackage{amsmath}
\usepackage{amsthm}
\usepackage{amsfonts}
\usepackage{amssymb}
\usepackage{color}
\usepackage{booktabs}
\usepackage[usenames,dvipsnames]{pstricks}
\usepackage{epsfig}
\usepackage{rotating,graphics,psfrag}
\usepackage{mathtools}
\usepackage[hyperfootnotes=false]{hyperref}
\usepackage{tikz}
\usepackage{pgfplots}
\usepackage{dsfont}
\usepackage{graphicx}%
\providecommand{\U}[1]{\protect\rule{.1in}{.1in}}
\IEEEoverridecommandlockouts
\usepgflibrary{decorations.shapes}
\usetikzlibrary{decorations.shapes, shapes, fit, arrows, positioning, trees, mindmap, calc}

\pgfplotsset{
  plot coordinates/math parser=false,
  log base 10 number format code/.code={
      $\displaystyle
      \ifthenelse{\equal{#1}{0.0}}
      {10^{\scriptscriptstyle\myphantom{\scriptscriptstyle-}{.3}\pgfmathprintnumber{#1}}\myphantom{\scriptscriptstyle-}{.7}}
      {10^{\scriptscriptstyle\pgfmathprintnumber{#1}}}
      $
  },
  every axis y label/.style={at={(ticklabel cs:0.5)},rotate=90,left=1mm,anchor=center},
  every axis x label/.style={at={(ticklabel cs:0.5)},below=1mm,anchor=center},
  every axis/.append style={
      footnotesize,
      width = .5\textwidth,
      legend style={font=\scriptsize,legend cell align = left,legend pos =north west},
      tick style={semithick},
  },
  every axis plot/.append style={thick},
}

\definecolor{darkgreen}{rgb}{0.12549019607843137255,0.4980392156862745098,0.16862745098039215686}
\definecolor{mycolor1}{rgb}{1.00000,0.00000,1.00000}%

\DeclareMathAlphabet{\mathbit}{OML}{cmr}{bx}{it}
\DeclareMathAlphabet{\mathsf}{OT1}{cmss}{m}{n}
\DeclareMathAlphabet{\mathbsf}{OT1}{cmss}{bx}{it}

\newcommand{\ve}[1]{\boldsymbol{#1}}
\newcommand{\F}[1]{\mathbb{F}_{#1}}

\DeclareMathOperator*{\argmin}{argmin}

\newcommand{\tblue}[1]{\textcolor{black}{#1}}

\title{\tblue{Towards Massive Connectivity Support for Scalable mMTC Communications in 5G networks}}

\author{\authorblockN{Carsten Bockelmann, Nuno K. Pratas, Gerhard Wunder, Stephan Saur, Monica Navarro, David Gregoratti, \\ 
Guillaume Vivier, Elisabeth de Carvalho, Yalei Ji, Cedomir Stefanovic, Petar Popovski, Qi Wang,\\
Malte Schellmann, Evangelos Kosmatos, Panagiotis Demestichas, Miruna Raceala-Motoc, Peter Jung, \\
Slawomir Stanczak and Armin Dekorsy}
}

\begin{document}
\maketitle

\begin{abstract}
 	\tblue{The fifth generation of cellular communication systems is foreseen to enable a multitude of new applications and use cases with very different requirements. A new 5G multi-service air interface needs to enhance broadband performance as well as provide new levels of reliability, latency and supported number of users. In this paper we focus on the massive Machine Type Communications (mMTC) service within a multi-service air interface. Specifically, we present an overview of different physical and medium access techniques to address the problem of a massive number of access attempts in mMTC and discuss the protocol performance of these solutions in a common evaluation framework.} 

\end{abstract}



\section{Introduction} 
\label{sec:introduction}

The prospect of billions of interconnected devices within the paradigm of the Internet of Things (IoT) has become one of the main drivers of the research and development in the ICT sector.
In fact, the 5G requirements for IMT-2020 include the support of a multiplicity of services and applications, with massive Machine Type Communications (mMTC) being one of the three cores services. The other core services being the Ultra Reliable Low Latency (URLLC) and the extreme Mobile Broadband (eMBB) communications.

The focus in this paper is on the massive access protocols and multi-user decoding techniques associated with the support of the mMTC core service. The objective is to attach a large number of low-rate low-power devices, termed Machine-Type Devices (MTDs), to the cellular network.
There are multiple factors that demand increased number of connected MTDs: the smart-grid, large scale environment and structure monitoring, asset and health monitoring, etc.
Typically, these MTDs connect asynchronously and sporadically to the network to transmit small data payloads. 
Connected objects include various \tblue{types} with an extremely wide set of requirements: for instance, a connected goggle providing augmented reality would require lower latency and higher throughput compared to a connected smoke detector. However, it is commonly understood that mMTC indicates the family of devices requiring sporadic access to the network to transmit small data payloads. The sporadic access leads to having an unknown, random subset of devices being active at a given transmission instant or frame, which necessitates the use of some form of random access protocol.

Most of the existing MTC connections, not necessarily massive, are wireless and take place via open standard short range technologies that operate in unlicensed spectrum, such as IEEE 802.15.x and 802.11.
Another trend is seen in the proprietary technologies for wide-area IoT, such as SIGFOX \cite{sigfox} and LoRA \cite{lora}, addressing the physical domain not covered by short-range technologies and thus providing a clear indication of an emerging market that is yet to be filled by service providers.
Until recently, the cellular standards could only provide access to MTDs via SMS or GPRS.
This approach suffers from coverage limitations (e.g., in deep indoor for instance for gas or water meters), non-optimized hardware and limited subscription models. Moreover, the 2G/3G systems were not designed to handle thousands of sporadically active MTDs. 
As a result, the 3GPP has extended the support of LTE to MTC with the standardization of cat-M and NB-IoT in 2015-2016. Those standards meet most of the mMTC requirements, but still need to be improved to support the massive number of terminals with low capabilities, sporadic activity patterns, and short packet transmissions.

One of the major obstacles for the proliferation of efficient cellular access for mMTC stems from the deficiencies of the access reservation procedure, a key building block of the cellular access networking. \tblue{Currently, the access reservation procedure is designed to enable connection establishment from a relatively low number of accessing devices. Additionally, each device has moderate to high data-rate requirements such that the overhead of current access protocols with multiple phases is relatively small. Both assumptions, the low number of devices as well as moderate to high data rates are in contradiction to mMTC needs. Thus, enhancement of the access reservation procedure for mMTC traffic has been in the focus of both the research community \cite{surveyRACHLTE} and standardization \cite{3GPPTR37.869}. However, there is a common understanding that the mMTC traffic requirements call for a more radical redesign of the cellular access \cite{6736746}.}

Indeed, 3GPP has recently concentrated its standardization efforts in this regard in four parallel tracks, which are (i) LTE for M2M (eMTC), focusing on the modification of LTE radio access network (RAN) for mMTC services and targeted at devices with reduced air-interface capabilities~\cite{3GPPRP1511862015}, (ii) narrow-band IoT (NB-IoT) which targets low-cost narrow-band devices with reduced functionalities~\cite{3GPPNBIOT2015}, (iii) extended coverage GSM for IoT (EC-GSM-IoT)~\cite{3GPPTS43.064} and (iv) the support of mMTC in 5G.  
In the efforts (i)-(iii), the goal can be summarized as \cite{3GPPTR22.368}: improved indoor coverage (15-20dB when compared to current cellular systems) and outdoor coverage up to 15~km, support of massive number of low data-rate devices with modest device complexity, improved power efficiency to ensure longer battery life, reduced access latency and efficient co-existence with the legacy cellular systems.
In (iv), the development towards 5G has started in 3GPP; and while the first phase, to be standardized in Release 15 \cite{Schaich2016_ETT}, focuses on extreme MBB (eMBB) services, the URLLC and mMTC will be in the focus of the following phases.

\tblue{In this paper we summarize several approaches to address the massive access problem for mMTC in 5G and present an evaluation framework to assess the performance of the presented approaches in terms of the access protocol performance. The presented solutions are part of the main innovations and outcomes of the FANTASTIC-5G project~\cite{F5GWebsite} \footnote{FANTASTIC-5G is the phase 1 project of Horizon 2020 in the framework of 5G PPP dealing with the air interface below 6GHz with time-line completing on the July 2017.}. First, we will outline the overall mMTC challenges and the specific research questions to be addressed in section~\ref{sec:challenges} (also see the overview paper \cite{Schaich2016_ETT}) and provide a short overview of the state of the art MTC systems in section~\ref{sec:mmtc_state_of_the_art} . Then our system level and evaluation approach will be outlined in section~\ref{sec:fantastic_5g_approach} and detailed technical approaches and their achieved performance for pure MAC protocols will be discussed in section~\ref{sub:mac_based_schemes} and combined PHY\& MAC approaches in section~\ref{sub:phy_and_mac_integrated_schemes}. Finally, we will present the results and compare different solutions in terms of their requirements and advantages. The paper wraps with conclusions in section~\ref{sec:conclusions}.}


\section{mMTC Challenges} 
\label{sec:challenges}
\begin{figure}[t]
	\centering
	\includegraphics[width=\linewidth]{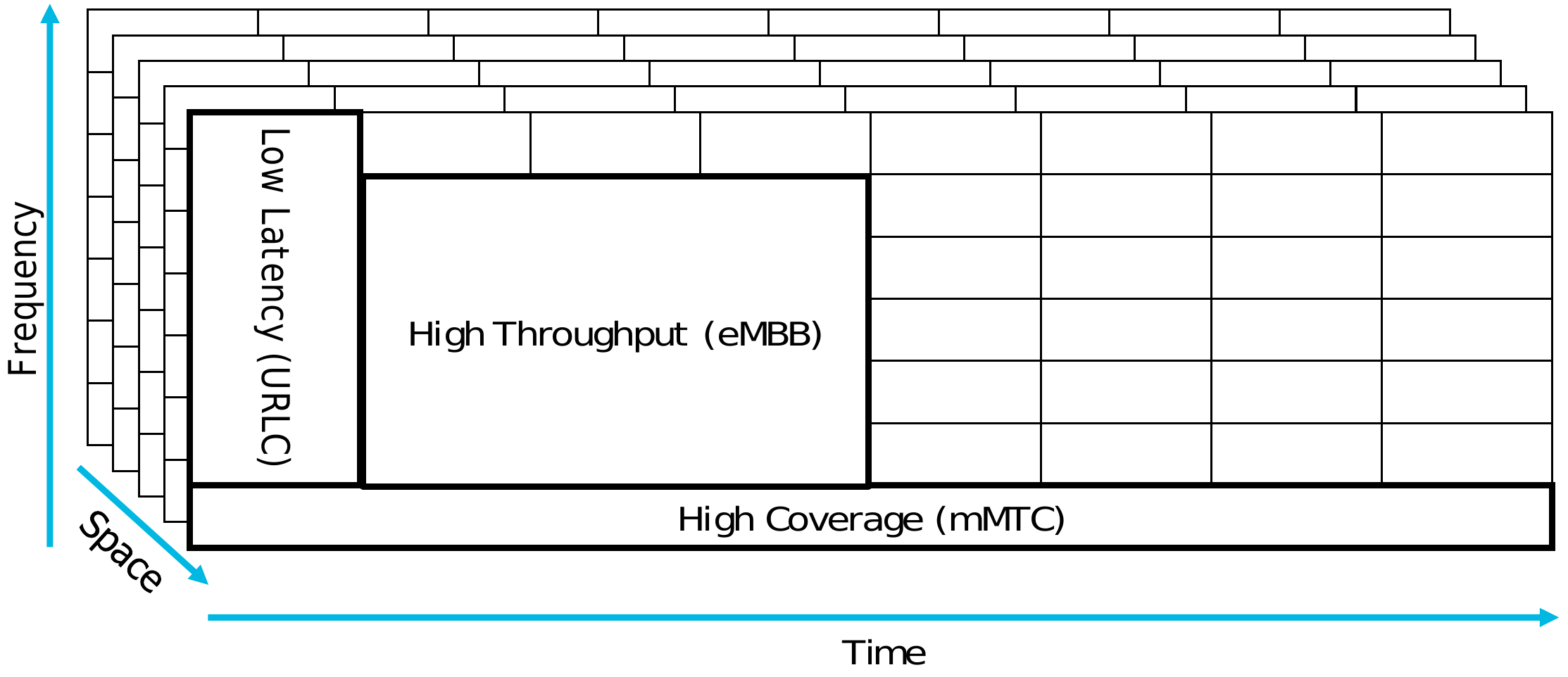}
	\caption{Multi service integration.}%
	\label{fig:MultiServiceIntegration}%
\end{figure}

Many MTC applications are already served by \tblue{today's} communication systems. However, the characteristic properties of mMTC, i.e.~the massive number of devices and the very short payload sizes, require novel approaches and concepts. 5G offers the opportunity to tackle the critical challenges in a seamless cellular system combining mMTC and all the other services.
This leads to the following mMTC challenges:
\begin{itemize}

\item \textbf{Control \tblue{signaling} challenge}:  In the existing LTE specification an endless cascade of signalling exchange between MTD, eNodeB and core network is initiated if an MTD is in idle mode and intends to send one single small packet. The overall number of sent bits is dominated by control information, and the actual data becomes negligible. Therefore, a 5G system must provide low overhead data transmission modes through novel MAC and PHY design. Additionally, higher layer enhancements, such \tblue{as} radio resource control \tblue{signaling}, are urgently required to lower the overhead on the reconnecting and re-authentication of idle users. Finally, methods that enable the transmission of small data packets over the control plane should be considered.

\item \textbf{Access capacity challenge}: In LTE the first step to accessing the system or reconnecting when the device in idle mode, is the access reservation protocol. The throughput of the LTE access reservation protocol is severely degraded since there is no specific collision resolution procedure in physical (PHY) and medium access (MAC) layer. A 5G system must at least enhance the access reservation protocol through novel MAC and PHY approaches to \tblue{support} a massive number of devices.

\item \textbf{Power consumption}: The MTDs are often battery powered and require 10+ years of autonomy. For that purpose, the access and communication schemes should be power efficient. This challenge is also related to the type of connection: either always UL triggered (mobile originated) or DL traffic (network originated communications) are considered. For instance, in a Sigfox model, communication is always triggered by an UL request, which helps in terms of power consumption (no need to wake up for paging channels).

\item \textbf{Multi-Service Integration}: LTE is mostly focused on MBB services and uses a single frame definition and common control channels for these services. In order to enable coexistence of services with very different requirements, 5G needs to include flexible frame definitions, a robust waveform and flexible control channel design to allow for dynamic bandwidth sharing and different PHY/MAC approaches. An example is provided in Fig.~\ref{fig:MultiServiceIntegration} on the multi-service integration over frequency, time and space resources.

\end{itemize}
\tblue{We focus on the MAC and PHY layer enhancements required to solve the outlined challenges. On the one hand, access protocols with novel waveforms are considered to enable spectral and temporal asynchronicities with very low control overhead; on the other hand, several MAC and PHY approaches and their combination are presented to specifically address the first two challenges "Control Signalling" and "Throughput". The focus of this paper is on summarizing potential solutions and providing insight on the access protocol throughput and latency of these solutions as will be discussed in section~\ref{sec:fantastic_5g_approach}. Novel waveforms are only exploited as enabling technology for a novel access protocol here, an exhaustive treatment of the waveforms being considered in a 5G setting is provided in~\cite{FANTASTIC-5GD3.1,FANTASTIC-5GD3.2}}.

\subsection{System Level Considerations} 
\label{sub:system_level_considerations}

When the scenarios under examination are extended in order to examine topologies with many \tblue{cells} and in order to take into consideration higher layer functionalities then several system level considerations emerge. In addition, system level scenarios may include cooperative functions between two or more cells (e.g. by using the X2 interface) such as coordinated power control and mobility. In such environments,  one can summarize the main system level topics of mMTC as: a) inter-cell interference from devices connected at neighboring cells; b) power control considerations; c) frame structure considerations and; d) intra-cell interference caused by asynchronous transmissions.

In scenarios in which several cells exists, interference \tblue{emerges} both among intra- and inter-cell devices. Regarding intra-cell interference, it emerge in cases of contentions, thus regarding the mMTC access protocols, interference emerges in the access notification stage of multi-stage and two-stage access protocols or during the combined access and data phase of one-stage protocols. On the contrary, inter-cell interference may emerge in any phase of the system including access, connection establishment and data phases, regardless of the selection of the access protocol.

In mMTC scenarios with a single cell, power control mechanisms are targeting to minimise the interference between devices (intra-cell interference) and in increasing the power efficiency to ensure longer battery life. In a multiple cell scenario, power control mechanisms are also targeting to minimise the inter-cell interference in addition to the above. In this direction, several coordinated power control mechanisms \tblue{exist} which study the trade-off between the effectiveness (preciseness of power control) and the overall overhead. 
Among the innovation of FANTASTIC-5G is the proposition of flexible frame definitions appropriate for a multi-service environment (Fig.~\ref{fig:MultiServiceIntegration}). In this direction, in contrast to eMBB services which are supported by numerologies with typical LTE TTIs (e.g 1ms) and URLLC service with strict latency requirements supported by small TTIs (e.g. 0.25ms), the special requirements of  mMTC services can be satisfied by numerologies with long TTIs and short subcarrier spacing in order to increase coverage and decrease device complexity and power consumption.

Regarding the loose uplink synchronization, one main limitation of the mMTC devices with sporadic uplink data is that they use the downlink channel for synchronization. This is not a major problem in small to medium sized cell environments (e.g., with inter-site distance 500m) and in cases of channel realizations with low delay spread values (e.g., EPA~\cite{3GPPTS36.104}), because in these cases the use of cyclic prefix (CP) compensates for any deviations of the transmission from the detection window reference time. But, in case of large cells (e.g. inter-site distance $>1500$m) and for channels with high delay spreads (e.g. ETU~\cite{3GPPTS36.104}), the deviation can become larger, especially for the devices afar from the base station, and can surpass the selected CP value. In this case, the transmission is considered asynchronous to the detection window and it produces interference to the transmissions adjacent in frequency. The power of this interference is affected by various parameters (e.g., the size of two bursts, the existence of guard bands between them, etc.). In FANTASTIC-5G a set of new waveforms are proposed with properties which can limit and in some cases eliminate the interference effects due to asynchronicity~\cite{FANTASTIC-5GD3.1,FANTASTIC-5GD3.2}.



\section{mMTC State of the Art} 
\label{sec:mmtc_state_of_the_art}

\tblue{Several ongoing efforts aim to support mMTC in commercial communication systems, but most of these only support parts of the mMTC requirements. Short payload packets and extended coverage are already available in some of the solutions. However, the problem of a massive number of devices attempting access has not been solved. In the following, we provide a short overview of mMTC systems currently available or under development covering 3GPP systems as well as Non-3GPP systems.}

\subsection{Non-3GPP \tblue{Low Power Wide Area Networks}} 
\label{sub:non_3gpp_approaches}

LoRA is a Low Power Wide Area Network (LPWAN) \tblue{and} is typically laid out in a star-of-stars topology in which gateways relay messages between end-devices and a central network server in the network back-end, \cite{lora}.
The communication between end-devices and gateways is spread out on different frequency channels and data rates. The selection of the data rate is a trade-off between communication range and message duration \tblue{offering a range of }0.3 kbps to 50 kbps through an adaptive data rate scheme. The access is based on a proprietary chirp based spread spectrum scheme and \tblue{the} MAC protocol is based on frequency and time ALOHA. LoRA operates in the sub-GHz bands and the vendors claim coverage on the order of 10–15 km in rural areas and 3–5 km in urban areas.

Sigfox is also a LPWAN that supports infrequent bi-directional communication, employs ultra narrow-band (UNB) wireless modulation as access technology, while the MAC protocol is based \tblue{on} frequency and time ALOHA~\cite{sigfox,7721743}.
The upper layers are proprietary and their definition is not public.
The vendor claims coverage on the order of 30–50 km in rural areas and 3–10 km in urban areas.

IEEE 802.11ah is a WAN, offering low-power and long-range operation. The operating frequency of IEEE 802.11ah is below 1GHz, allowing a single access point (AP) to provide service to \tblue{an} area of up to 1 km.
The PHY and MAC protocol operation is similar to the one present in the 802.11 family of protocols, extended with the introduction of restricted access window during which only certain number of devices are allowed to contend based on their device IDs~\cite{7312879}.

There are other network systems built on top of IEEE 802.15.4 (6LoWPAN, ISA100.11a, WirelessHart) which are focused on low number of devices while providing reliability guarantees.
Finally, there are other network systems with their own protocol stack such as Ingenu and Weightless.

All of the \tblue{presented LPWANs} assume a rather simple physical layer processing and are not capable of coping with massive number of simultaneously active devices.


\subsection{3GPP \tblue{Low Power Wide Area Networks}} 
\label{sub:3gpp_lpwan}

Until recently, MTDs were being served by 2G based solutions. However, with the success of non-3GPP technologies as previously described, such as LoRA, Sigfox, Ingenu \tblue{and} Weightless, the cellular industry decided to accelerate the definition of an efficient MTC set of solutions and came up with solutions standardized in 2016.
The aim was to introduce new features to the LTE releases that would support IoT-like devices and would exploit the existing 4G coverage around the world. However, these new features would need to align with the new IoT key requirements which can be summarized as following:
\begin{itemize}
	\item Low cost receiver devices ($~2-5$ \$);
	\item Long battery life ($>10$ years)
	\item Extended coverage ($+15$~dB) \tblue{over LTE-A}
\end{itemize}

In order to achieve the three objectives (cost, power efficiency, extended coverage), design choices were made:
\begin{itemize}
	\item Single antenna design (to reduce cost)
	\item Half duplex transmission (to reduce cost)
	\item Narrow band reception (to reduce cost, power consumption)
	\item Peak rate reduction (to reduce cost, complexity)
	\item Limited MCS and limited number of Transmit modes (to reduce complexity)
	\item Lower transmit power (to reduce power consumption)
	\item Extended DRX and new power saving modes (to reduce power consumption)
	\item Transmission repetition (for enhanced coverage)
\end{itemize}
Three types of IoT devices are currently supported in the 3GPP standards up to Release 13.
These are the category M1 (Cat-M1), NB-IoT (NB1) and the extended coverage GSM (EC-GSM). The latter solution targets a very specific market (2G only) and is most likely to stay as a niche technology as the 2G systems spectrum resources are re-farmed into 4G. 

\subsubsection{Cat-M1} 
\label{sub:emtc}

The eMTC (now denoted as cat-M1) comes from the need to support simpler devices than the UE types defined currently, while being capable to take advantage of the existing LTE capabilities and network support.
The changes in comparison with the LTE system take place both at the device and at the network infrastructure level, where the most important one is the reduction of the device-supported bandwidth from $20$~MHz to $1.4$~MHz in both downlink and uplink~\cite{3GPPRP-151186}.
The main consequence of this change is that the control signals (e.g. synchronization or broadcast of system block information) which are currently spread over the $20$~MHz band, will be altered to support the coexistence of both LTE-M UEs and the standard, more capable, UEs.
Another important feature of this new UE category is the reduced power consumption, achieved by the transceiver-chain complexity and cost reduction, such as support of uplink and downlink rate of $1$ Mbps, half-duplex operation, use of a single antenna, reduced operation bandwidth of $1.4$~MHz, and reduction of the allowed maximum transmission power from $23$~dBm to $20$~dBm.
Furthermore, there is the requirement to increase the cellular coverage of these LTE-M UEs by providing up to $15$~dBs extra in the cellular link budget.

The preamble structure and access procedure are the same as in LTE, with the introduction of a simplified procedure without the security overhead. It is focused on increasing coverage, while still keeping LTE-like functionality. 


\subsubsection{NB1} 
\label{sub:nb_iot}

The NB-IoT (also denoted as NB1) pertains to a clean slate design of an access network dedicated to serve a massive number of low throughput, delay tolerant and ultra-low cost devices.
NB-IoT can be seen as an evolution of eMTC in respect to the optimization of the trade-off between device cost and capabilities; as well as a substitute to legacy GPRS to serve low rate IoT applications.
The main technical features are: (i) reduced bandwidth of 180~kHz in downlink and uplink; (ii) maximum device transmission power of $23$~dBm; and (iii) increased link budget by $20$~dB extra when compared with commercially available legacy GPRS, specifically to improve the coverage of indoor IoT devices.
This coverage enhancement can be achieved by power boosting of the data and control signals, message repetitions and relaxed performance requirements, e.g., by allowing longer signal acquisition time and higher error rate.
An important enabler for this coverage enhancement is the introduction of multiple coverage classes, which allow the network to adapt to the device's coverage impairments. 

It has a new PRACH structure based on multi-hopping \tblue{and is} not based on Zadoff-Chu sequences \tblue{like in LTE}. There are three versions of the access protocol (full similar to LTE, medium similar to the optimized \tblue{access} in eMTC and light with a preamble followed by data transmission). The main focus of the NB-IoT is on providing extreme coverage\tblue{, with supported number of users similar to LTE-M~\cite{lauridsen2016}}.



\section{\tblue{mMTC in a Multi-Service Air Interface}} 
\label{sec:fantastic_5g_approach}

\subsection{System Model and Assumptions}
\label{sec:system_model}

\tblue{In general, we assume mMTC to be part of} a multi-service air interface suitable to serve all services envisioned in 5G in a single air interface~\cite{ETT:ETT3050}, as depicted in Fig.~\ref{fig:MultiServiceIntegration}.
The base physical layer assumption \tblue{for such a multi-service air interface is a multi-carrier system with a suitable waveform and flexible numerologies as standardized for New Radio (NR) in 3GPPP}. Thus, the mMTC service (denoted as MMC in FANTASTIC-5G) may use part of a resource block grid as it is depicted in Fig.~\ref{fig:MultiServiceIntegration} and can be organized using all or part of these resources. Of course, the amount of resources available for mMTC and the numerology used will vary according to higher layer management functionalities that balance service requirements in a given scenario or cell. For example, LTE provides only limited resources for the PRACH that facilitates the access reservation protocol in LTE and thereby limits the number of serviceable users.
\begin{table}
	\centering
	\begin{tabular}{llp{4.3cm}}\hline
	Parameter & Value & Explanation \\
	\hline
	TTI & 1~ms & \\
	Bandwidth & 10~MHz & 50 PRBs per TTI \\
	Allocation size & 1 PRB & 1 PRB = 1~ms x 180~kHz\\
	\tblue{No.~of antennas} & \tblue{1} & \tblue{Base assumption is single antenna at UE and BS}\\
	\hline
	Traffic model & Poisson & Arrival rate $\lambda$ \\
	Packet size & 8 Bytes & \\
	\hline
	\parbox[t]{1.5cm}{Mean \\waiting time} & 0.5~ms & Avg.~time offset between wake-up of the UE and the beginning of the next TTI when a SR is sent \\
	\parbox[t]{1.6cm}{ACK/NACK \\response time} & 3~ms & A request or packet sent in TTI $i$ is followed by ACK/NACK at TTI $i+3$, earliest retransmission then is TTI $i+4$ \\
	\parbox[t]{1.5cm}{Random\\ Back-off} &  0..10~ms & Uniform distribution, back-off after NACK \\
	\parbox[t]{1.5cm}{Max \\Retransmissions} & 4 & The fourth NACK is the "final" NACK\\
	\hline
	\end{tabular}
	\caption{Basic assumptions for MMC evaluation}
	\label{tab:params}
\end{table}

\tblue{In contrast to this general view on a multi-service architecture, we aim to present different solutions in a comparable framework such that the access protocol performance can be gauged by different key performance indicators (KPIs). Thus, for evaluation of the proposal described in section~\ref{sub:mac_based_schemes} and \ref{sub:phy_and_mac_integrated_schemes}}, we consider a single cell scenario using the basic PHY layer assumptions summarized in Table~\ref{tab:params}. This allows the evaluation of the base performance of different MMC PHY/MAC concepts. Furthermore, we assume a generic OFDM waveform \tblue{as base assumption} that excludes topics like synchronization robustness or service separation solved by appropriate waveform choices~\cite{FANTASTIC-5GD3.2}.

\subsection{Building Blocks} 
\label{sec:building_blocks} 

\tblue{In order to address the mMTC challenges we have identified a number of building blocks that are classified into (i) Physical Layer, (ii) MAC layer, (iii) RRC layer and (iv) Waveforms. The focus of this paper is on the first two, i.e.~the physical and MAC layers. However, PHY and MAC enhancements alone will not be able to solve the massive access challenges. Therefore, we also provide a short outlook on RRC and waveforms}.

\subsubsection{Physical Layer} 
\label{sub:physical_layer}

The design of access reservation protocols is usually based on idealized assumptions about the PHY performance and behavior. A classical assumption in contention based protocols is that concurrently active users are colliding and cannot be retrieved. \tblue{Recently,} MAC protocol analysis took the capture effect~\cite{6155698} into account, i.e.~the decodability of users with sufficiently different powers such that at least one can be still decoded. PHY layer technologies that are able to resolve more collisions through advanced receiver processing like successive interference cancellation (SIC) have been in focus to enhance the performance of the overall access protocol. Furthermore, the performance of such technologies in different fading scenarios as well as under the assumption of asynchronous communication, strongly determines the performance baseline of all MAC protocols based on specific PHY solutions. \tblue{In FANTASTIC-5G we studied different PHY collision resolution techniques in combination with various access protocols}. On the one hand classical multi-user detection (MUD) as well as Compressive Sensing based enhancements are considered\tblue{,} and on the other hand also Compress- or Compute-and-Forward based schemes are considered that can be closely related to or even combined with advanced protocols like Coded Random Access.


\subsubsection{Medium Access Control Layer} 
\label{sub:medium_access_layer}

\begin{figure}[t]
	\centering
	\includegraphics[width=\linewidth]{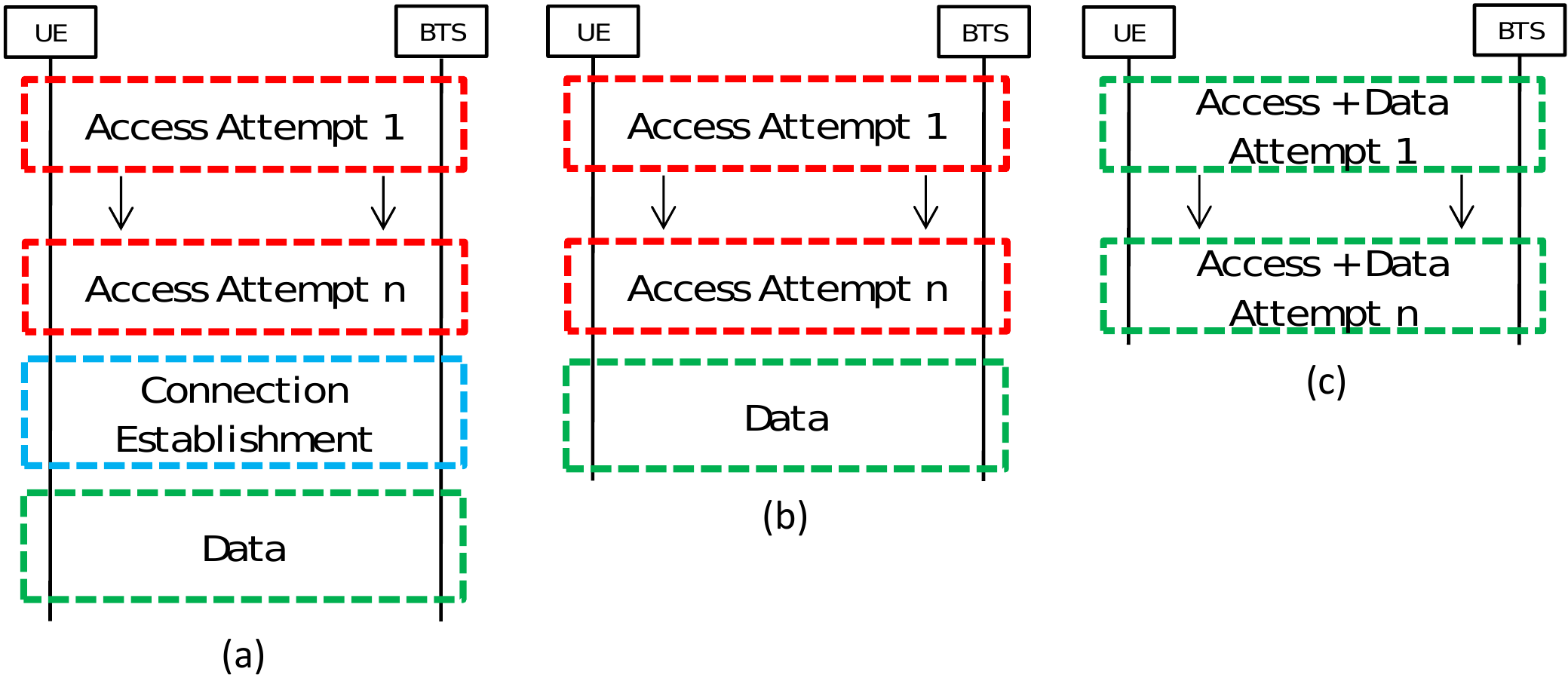}
	\caption{High level description of the three \tblue{considered access protocols} types: (a) Multi-stage access protocol with an access, connection
	establishment and data phase; (b) Two-stages access protocol with access and
	data phases; and (c) One-stage access with combined access and data phase.}%
	\label{fig:AccessProtocolClassification}%
\end{figure}

\tblue{We} distinguishes three types of access protocols: (a) multi-stage;
(b) two-stage; and (c) one-stage. These can be interpreted very
differently, and each \tblue{of the three types may} contain several access protocol variants. We
depict these in Fig.~\ref{fig:AccessProtocolClassification}.
\tblue{A multi-stage access protocol (a), for which the current LTE connection
establishment protocol is a prime example, is composed of at least three
phases, the access, connection establishment (including authentication and
security) and finally the data phase. A two-stage access protocol (b) allows
the UE separating the access notification stage with its data delivery stage, 
e.g. through an intermediate feedback message. This leaves room for feedback 
and resource allocation to the UE from the eNB. The feedback could be power
control and timing alignment. What is meant by a one-stage access protocol (c) is
that both the access notification and data delivery need to be done in a single
transaction, e.g. using one or several consecutive packets or in a single transmission. 
All three types of access protocols can lead to scheduled access mode, where the devices after
establishing a connection to the network do not need to re-establish access
in future attempts.}

The work presented in this paper focuses on two-stage and one-stage access protocols.
As a common assumption the \tblue{signaling} associated with the first connection establishment
(mostly the establishment of mutual authentication and security) is assumed to be reused
from a previous session where some incarnation of a full multi-stage access protocol took place. 


\subsubsection{Radio Resource Control Layer} 
\label{sub:radio_resource_control_layer}

\tblue{A major observation beyond PHY and MAY layer was that the transition from idle mode to connected mode and vice versa used in today's systems must be simplified or even avoided.} Connectionless transmission of small packets from UEs once registered and authenticated in the network may reduce the required number of signalling messages significantly. In this case, a small packet must comprise both source and destination addresses and payload. An important component in the reduction of the required signalling, upon connection establishment, is the addition of new RRC states such as the RRC extant state (see~\cite{FANTASTIC-5GD4.2}) which will allow the devices to maintain the security context active over a long period. 


\subsubsection{Waveforms} 
\label{sub:waveforms}

Another major conclusion is that due to properties of \tblue{new 5G waveforms tight uplink synchronization is not required anymore for the small packets typical in mMTC (see~\cite{FANTASTIC-5GD3.1,FANTASTIC-5GD3.2})}. This allows to compress or even avoid broadcast messages that are usually required for synchronous operation. One example is the random access response (RAR) in LTE which consists of 56 bits for each UE that has sent a detected preamble. Essentially, RAR comprises a temporary identifier, a time offset value, and a grant for the subsequent \tblue{signaling} messages. While for the one-stage access protocols discussed here the RAR can be completely omitted, others like the contention-based two-stage variants combined with new waveforms can significantly reduce such overhead.



\subsection{Key Performance Metrics} 
\label{sub:key_performance_metrics}

\tblue{To evaluate the performance of our various contributions on the PHY and MAY layers detailed in sections~\ref{sub:mac_based_schemes} and \ref{sub:phy_and_mac_integrated_schemes} we consider two key performance indicators:} 
\begin{itemize}
\item The \textbf{Protocol Throughput} (TP) denotes the total number of served
devices per \tblue{TTI. It directly addresses the massive access problem by showing how many users can be served given a certain access load.}

\item The \textbf{Access Latency} (AL) measures the amount of time (measured in
TTIs) between \tblue{$T_{1}$} the time instant when a device has new data to
transmit (packet arrival at the device) and \tblue{$T_{2}$} the time instant when
the device's data is received successfully (packet arrival at
the receiver, which in most cases is the Base Station). \tblue{Here, it complements the throughput to provide a complete view. Without latency considerations the throughput could be arbitrarily enhanced by aggregation of access opportunities and longer back-off times. Therefore, technologies can only be fairly compared if both KPIs are considered together.}
\end{itemize}

\tblue{This manuscript gives a compact overview of the proposed protocols and highlights evaluation results obtained in the EU funded project FANTASTIC-5G. More details on the evaluation of the proposed protocols and additional results can be found~\cite{FANTASTIC-5GD4.2}.}



\section{\tblue{MAC Protocol Procedures}} 
\label{sub:mac_based_schemes}
\tblue{In this section we present three different MAC layer approaches using idealized models of the physical layer. First, we present results for One-Stage vs Two-Stages Access Protocols (OSTSAP) with different number of preambles and additionally exploiting decoding of multiple collisions (capture effect) showing that one-stage protocols offer better latency whereas two-stage protocols with collision resolution allow for much higher throughput. Second, we present Signature based Access with Integrated Authentication (SBAIA) that extends the idea of random access preambles like in standard LTE to a signature formed of multiple preambles enabling much higher throughput with added functionality like authentication. Finally, we present Non-Orthogonal Access with Time-Alignment Free Transmission (NOTAFT) that exploits the relaxed timing constraints of Pulse-shaped OFDM and MIMO processing to lower the signaling overhead for MTDs and enable massive access.}

\subsection{One-Stage vs Two-Stages Access Protocols \tblue{(OSTSAP)}} 
\label{sub:one_stage_vs_two_stages_access_protocols}
In this section we describe implementation variants of the generic two-stage and one-stage schemes shown in Fig. 2(b) and (c), respectively. In contrast to the other solutions presented in this paper, the performance evaluation is limited to pure protocol performance. In case of single-user detection (SUD), this means that two packets collided on the same data resource are always lost, whereas a single packet is always successful. In case of multi-user detection (MUD), we apply an idealized model to get the upper bound of the potential performance gain~\cite{StephanSaur2017}. We assume that at most two superimposed packets on the same data resource can be decoded given that the UEs have utilized different preambles. Unpredicted overlapping of more than two packets leads to the loss of all of them. Of course, this scheme can be easily extended to more than two users. In a more general view, the probability of successful decoding of any packet $P(n)$, given that $n$ packets overlap, depends on multiple parameters, e.g., the distribution of receive power at the BTS, the modulation and coding scheme (MCS) and the multiple access method on the PHY layer itself~\cite{6155698}.

A detailed introduction of the protocol options can be found in~\cite{7421049}. The following paragraphs briefly summarize the two-stage protocol in Fig.~\ref{fig:AccessProtocolClassification}(b): The UE sends a random preamble sequence, also referred to as service request. We assume a set of $S$ sequences which can be uniquely detected and separated at the BTS through a correlation receiver. However, the BTS cannot distinguish whether just one single UE or several UEs have sent the same sequence. The latter case is referred to as preamble collision. With increasing $S$, this probability can be reduced at cost of a larger amount of required radio resources $M_S$. Given a constant number of resource units per time slot, $M = M_S + M_D$, increasing $M_S$ reduces the available data resources $M_D$ accordingly. Without loss of generality we assume in the following a preamble signal generation and transmission scheme equivalent to the Physical Random Access Channel (PRACH) in LTE, and resource units mimic a Physical Resource Blocks (PRB) stacked in frequency dimension.

The BTS broadcasts information related to the assignment of radio resources. In the simplest case this is a binary vector $V$ of length $S$ indicating whether or not the originator of the respective sequence is allowed to transmit its data packet in the second stage. This implies a fixed mapping between preamble sequence and data resource. Typically, the number of sequences $S$ exceeds the number of data resources $M_D$ by an over-provisioning factor $N$, i.e.~$S = N M_D$. Consequently, $N$ sequences point to one single data resource. The BTS without multi-user capability will therefore acknowledge just one detected preamble and reject the remaining. In case of MUD, a second detected preamble is acknowledged as well. A more sophisticated feedback scheme comprises a resource index instead of just one bit ACK/NACK, allowing the BTS for a fully flexible assignment of the detected service requests to the available data resources at cost of a larger downlink signaling overhead. A further enhanced scheme includes additionally the queue length of waiting UEs that could not yet be served. This enables a distribution of the detected service requests in both frequency and time domain, i.e.~surplus service requests are automatically shifted to the next free time slot. In the second stage the acknowledged data packet transmission takes place. In case of any error, the retransmission scheme with parameters in Table~\ref{tab:params} is initiated.

In the one-stage protocol shown in Fig.~\ref{fig:AccessProtocolClassification}(c), the intermediate feedback after preamble detection is missing. The main advantage is the acceleration of the complete process. Preamble for activity detection and data packet can be transmitted in the same time slot. However, the capability of the two-stage protocol to control data packet transmissions and to reduce collisions is not present any more. It is therefore straightforward to combine one-stage access with MUD. A significantly high over-provisioning factor $N$ allows the BTS to separate the service requests and to gain awareness how many data packets overlap on each of the $M_D$ resources.
\begin{figure}[ht!]
	\centering
%
%
%
%

\begin{tikzpicture}
\begin{axis}[%
small,
width=0.8\columnwidth,
height= 4.5cm,
scale only axis,
  try min ticks=6,
  max space between ticks=30pt,
  x tick label style={
    /pgf/number format/.cd,
    fixed,
    precision=2
  },
xmajorgrids,ymajorgrids,
xmin=0, xmax=2*50,
xlabel={Arrival Rate $\lambda$ [arrivals/TTI]},
ymin=0, ymax = 50,
ylabel={Protocol Throughput [users/TTI]},
label style ={font=\footnotesize},
legend cell align={left},
legend style={at={(1.075,1)}, anchor=north east,font=\tiny,text height=1ex,inner xsep=1pt,inner ysep=0pt,nodes={inner sep=0pt,text depth=0.15em},}]]

\addplot [smooth,
mark=x,
mark repeat=5,
color=blue,
solid,
x filter/.code={\pgfmathparse{\pgfmathresult*38}\pgfmathresult}]
table{
     0.1000    5.0074
    0.2000    9.8885
    0.3000   13.5457
    0.4000   12.5411
    0.5000    9.0347
    0.6000    6.1562
    0.7000    4.1643
    0.8000    2.8029
    0.9000    1.9049
    1.0000    1.2625
    1.1000    0.8706
    1.2000    0.6012
    1.3000    0.4060
    1.4000    0.2919
    1.5000    0.1905
    1.6000    0.1364
    1.7000    0.0972
    1.8000    0.0652
    1.9000    0.0460
    2.0000    0.0346
};
\addlegendentry{One-Stage (SUD) with 108 preambles}

\addplot [smooth,
mark=square,
mark repeat=5,
color=red,
solid,
x filter/.code={\pgfmathparse{\pgfmathresult*50}\pgfmathresult}]
table{
    0.1000    5.0102
    0.2000   10.0294
    0.3000   14.9611
    0.4000   19.7559
    0.5000   23.9182
    0.6000   23.4558
    0.7000   17.0267
    0.8000   11.5940
    0.9000    8.0121
    1.0000    5.4639
    1.1000    3.8095
    1.2000    2.6401
    1.3000    1.8617
    1.4000    1.3094
    1.5000    0.9355
    1.6000    0.6500
    1.7000    0.4869
    1.8000    0.3285
    1.9000    0.2480
    2.0000    0.1792
};
\addlegendentry{One-Stage (MUD) with 108 preambles}

\addplot [smooth,
mark=o,
mark repeat=5,
mark options={solid},
color=red,
dashed,
x filter/.code={\pgfmathparse{\pgfmathresult*50}\pgfmathresult}]
table{
    0.1000    5.0028
    0.2000    9.9803
    0.3000   15.0002
    0.4000   19.9569
    0.5000   24.7791
    0.6000   28.2260
    0.7000   23.3328
    0.8000   15.3514
    0.9000   10.0056
    1.0000    6.7195
    1.1000    4.5812
    1.2000    3.1085
    1.3000    2.0846
    1.4000    1.4031
    1.5000    0.9676
    1.6000    0.6520
    1.7000    0.4575
    1.8000    0.3006
    1.9000    0.2101
    2.0000    0.1317
};
\addlegendentry{One-Stage (MUD) with 216 preambles}

\addplot [smooth,
mark=x,
mark repeat=5,
color=darkgreen,
solid,
x filter/.code={\pgfmathparse{\pgfmathresult*50}\pgfmathresult}]
table{
    0.1000    4.9791
    0.2000    9.9999
    0.3000   15.0121
    0.4000   19.9210
    0.5000   24.8517
    0.6000   28.8025
    0.7000   23.5024
    0.8000   20.1023
    0.9000   17.4006
    1.0000   15.1202
    1.1000   13.1184
    1.2000   11.4550
    1.3000    9.9454
    1.4000    8.6679
    1.5000    7.5160
    1.6000    6.5776
    1.7000    5.6904
    1.8000    4.9155
    1.9000    4.2369
    2.0000    3.6964
};
\addlegendentry{Two-Stage (SUD) with 108 preambles}

\addplot [smooth,
mark=star,
mark repeat=5,
mark options=solid,
color=darkgreen,
dashed,
x filter/.code={\pgfmathparse{\pgfmathresult*50}\pgfmathresult}]
table{
    0.1000    4.9949
    0.2000   10.0198
    0.3000   14.9989
    0.4000   20.0311
    0.5000   24.9533
    0.6000   29.9634
    0.7000   32.4352
    0.8000   29.6444
    0.9000   27.6570
    1.0000   25.9678
    1.1000   24.4011
    1.2000   22.9192
    1.3000   21.6131
    1.4000   20.3093
    1.5000   19.1325
    1.6000   17.9247
    1.7000   16.8414
    1.8000   15.8636
    1.9000   14.8787
    2.0000   13.9962
};
\addlegendentry{Two-Stage (SUD) with 216 preambles}

\addplot [smooth,
mark=square,
mark repeat=5,
color=black,
solid,
x filter/.code={\pgfmathparse{\pgfmathresult*50}\pgfmathresult}]
table{
    0.1000    5.0254
    0.2000   10.0461
    0.3000   15.0390
    0.4000   19.9700
    0.5000   24.8781
    0.6000   29.5035
    0.7000   30.4629
    0.8000   22.0162
    0.9000   15.2941
    1.0000   11.1359
    1.1000    8.3029
    1.2000    6.2116
    1.3000    4.6915
    1.4000    3.5757
    1.5000    2.7025
    1.6000    2.0577
    1.7000    1.5439
    1.8000    1.1866
    1.9000    0.8939
    2.0000    0.6719
};
\addlegendentry{Two-Stage (MUD) with 108 preambles}

\addplot [smooth,
mark=o,
mark repeat=5,
mark options=solid,
color=black,
dashed,
x filter/.code={\pgfmathparse{\pgfmathresult*50}\pgfmathresult}]
table{
    0.1000    5.0251
    0.2000    9.9607
    0.3000   15.0131
    0.4000   20.0183
    0.5000   24.9817
    0.6000   30.0387
    0.7000   34.9167
    0.8000   39.7598
    0.9000   44.4578
    1.0000   46.9909
    1.1000   35.5169
    1.2000   29.8424
    1.3000   25.5281
    1.4000   22.0064
    1.5000   19.2107
    1.6000   16.7734
    1.7000   14.6493
    1.8000   12.8635
    1.9000   11.2899
    2.0000    9.8499
};
\addlegendentry{Two-Stage (MUD) with 216 preambles}

\end{axis}
\end{tikzpicture}
	\caption{Protocol throughput of the one-stage and two-stage variants with SUD and MUD depending on the number of preambles S as a function of the arrival rate $\lambda$.}
	\label{fig:FigxProtocol_throughput}
	\vspace{-3pt}
\end{figure}
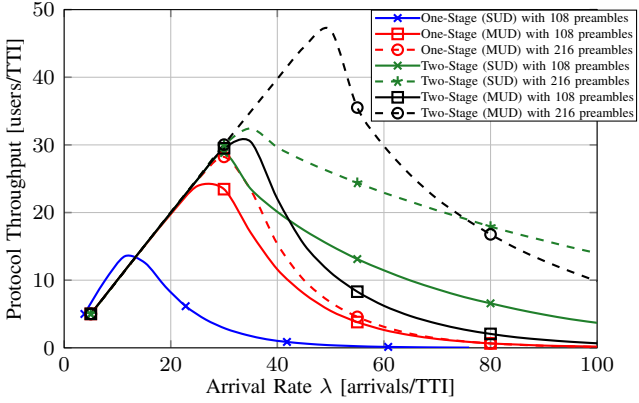

In Figure~\ref{fig:FigxProtocol_throughput} the achievable protocol throughput with SUD and MUD is depicted as a function of the arrival rate $\lambda$ for different large sets of preamble sequences $S$. Obviously, the two-stage protocol outperforms its counterpart with respect to throughput. Main reason for this result is the possibility to assign the available data resources through the intermediate feedback after preamble detection, and consequently to reduce collisions. With a larger set of preambles $S$, the performance can be significantly improved, especially in high load situations \tblue{(arrival rate $\lambda > 30$)}, and motivates further efforts to optimize the preamble sequence design for 5G. MUD improves the performance for both the one-stage and the two-stage access protocol because the number of resources is virtually increased. We remark that only cases with the same over-provisioning factor $N$, i.e.~the same ratio of available preambles and data resources can be directly compared, e.g., two-stage SUD with 108 preambles (solid dark green) and MUD with 216 preambles (dotted light green).
\begin{figure}[ht!]
	\centering
%
%
%
%

\begin{tikzpicture}
\definecolor{darkgreen}{rgb}{0.12549019607843137255,0.4980392156862745098,0.16862745098039215686}
\begin{axis}[%
small,
width=0.8\columnwidth,
height= 3.5cm,
scale only axis,
  try min ticks=6,
  max space between ticks=30pt,
  x tick label style={
    /pgf/number format/.cd,
    fixed,
    precision=2
  },
  y tick label style={
    /pgf/number format/.cd,
    fixed,
    precision=0
  },
xmajorgrids,ymajorgrids,
xmin=0, xmax=2*50,
xlabel={Arrival Rate $\lambda$ [arrivals/TTI]},
ymin=0, ymax = 25,
ylabel={Access Latency [ms]},
label style ={font=\footnotesize},
legend cell align={left},
legend style={at={(1,0)}, anchor=south east,font=\tiny,text height=1ex,inner xsep=1pt,inner ysep=0pt,nodes={inner sep=0pt,text depth=0.15em},}]]

\addplot [mark=x,
mark repeat=5,
color=blue,
solid,
x filter/.code={\pgfmathparse{\pgfmathresult*50}\pgfmathresult}]
table{
    0.1000    3.0017
    0.2000    5.4294
    0.3000    8.9883
    0.4000   12.2589
    0.5000   13.7303
    0.6000   14.3540
    0.7000   14.6488
    0.8000   14.7479
    0.9000   14.8984
    1.0000   14.9580
    1.1000   14.9840
    1.2000   14.9990
    1.3000   14.9971
    1.4000   14.9208
    1.5000   15.1541
    1.6000   14.8024
    1.7000   15.2230
    1.8000   14.3558
    1.9000   14.7074
    2.0000   14.7108
};
\addlegendentry{One-Stage (SUD) with 108 preambles}

\addplot [mark=square,
mark repeat=5,
color=red,
solid,
x filter/.code={\pgfmathparse{\pgfmathresult*50}\pgfmathresult}]
table{
    0.1000    1.9856
    0.2000    2.6843
    0.3000    3.6449
    0.4000    5.0590
    0.5000    7.4027
    0.6000   10.8151
    0.7000   13.1472
    0.8000   14.0957
    0.9000   14.4658
    1.0000   14.6229
    1.1000   14.8195
    1.2000   14.8714
    1.3000   14.9855
    1.4000   14.9678
    1.5000   14.9273
    1.6000   15.1047
    1.7000   14.9286
    1.8000   14.9926
    1.9000   15.0718
    2.0000   15.1698
};
\addlegendentry{One-Stage (MUD) with 108 preambles}

\addplot [mark=o,
mark repeat=5,
mark options={solid},
color=red,
dashed,
x filter/.code={\pgfmathparse{\pgfmathresult*50}\pgfmathresult}]
table{
 
    0.1000    1.7709
    0.2000    2.1904
    0.3000    2.7910
    0.4000    3.7504
    0.5000    5.2895
    0.6000    7.9596
    0.7000   11.9199
    0.8000   13.6239
    0.9000   14.2970
    1.0000   14.5746
    1.1000   14.7110
    1.2000   14.8500
    1.3000   14.9384
    1.4000   14.9143
    1.5000   14.9529
    1.6000   14.8844
    1.7000   14.9930
    1.8000   15.0986
    1.9000   15.0523
    2.0000   14.9889
};
\addlegendentry{One-Stage (MUD) with 216 preambles}

\addplot [mark=x,
mark repeat=5,
color=darkgreen,
solid,
x filter/.code={\pgfmathparse{\pgfmathresult*50}\pgfmathresult}]
table{
    0.1000    6.1335
    0.2000    6.9197
    0.3000    7.7727
    0.4000    8.7880
    0.5000   10.1024
    0.6000   13.1775
    0.7000   18.0814
    0.8000   19.1978
    0.9000   19.7938
    1.0000   20.1268
    1.1000   20.3606
    1.2000   20.5777
    1.3000   20.6543
    1.4000   20.7064
    1.5000   20.7851
    1.6000   20.8872
    1.7000   20.9483
    1.8000   21.0237
    1.9000   20.9535
    2.0000   20.9305
};
\addlegendentry{Two-Stage (SUD) with 108 preambles}

\addplot [mark=star,
mark repeat=5,
mark options=solid,
color=darkgreen,
dashed,
x filter/.code={\pgfmathparse{\pgfmathresult*50}\pgfmathresult}]
table{
    0.1000    5.8175
    0.2000    6.1546
    0.3000    6.4980
    0.4000    6.9302
    0.5000    7.3125
    0.6000    8.0043
    0.7000   13.2680
    0.8000   16.3600
    0.9000   17.4332
    1.0000   18.0404
    1.1000   18.4582
    1.2000   18.7011
    1.3000   18.9289
    1.4000   19.1135
    1.5000   19.2146
    1.6000   19.3257
    1.7000   19.4720
    1.8000   19.5319
    1.9000   19.5689
    2.0000   19.6287
};
\addlegendentry{Two-Stage (SUD) with 216 preambles}

\addplot [mark=square,
mark repeat=5,
color=black,
solid,
x filter/.code={\pgfmathparse{\pgfmathresult*50}\pgfmathresult}]
table{
    0.1000    6.1373
    0.2000    6.9035
    0.3000    7.7916
    0.4000    8.7955
    0.5000   10.0892
    0.6000   11.9513
    0.7000   17.1697
    0.8000   21.6751
    0.9000   22.7891
    1.0000   23.1992
    1.1000   23.2513
    1.2000   23.3072
    1.3000   23.3425
    1.4000   23.3138
    1.5000   23.3371
    1.6000   23.3709
    1.7000   23.2231
    1.8000   23.3177
    1.9000   23.2600
    2.0000   23.2978
};
\addlegendentry{Two-Stage (MUD) with 108 preambles}

\addplot [mark=o,
mark repeat=5,
mark options=solid,
color=black,
dashed,
x filter/.code={\pgfmathparse{\pgfmathresult*50}\pgfmathresult}]
table{
    0.1000    5.8117
    0.2000    6.1653
    0.3000    6.5152
    0.4000    6.9059
    0.5000    7.3530
    0.6000    7.8278
    0.7000    8.5587
    0.8000    9.7638
    0.9000   11.5961
    1.0000   14.6003
    1.1000   18.9084
    1.2000   19.8019
    1.3000   20.2673
    1.4000   20.5284
    1.5000   20.7184
    1.6000   20.8639
    1.7000   20.9657
    1.8000   21.0036
    1.9000   21.0828
    2.0000   21.0839
};
\addlegendentry{Two-Stage (MUD) with 216 preambles}

\end{axis}
\end{tikzpicture}
	\caption{Access latency of the one-stage and two-stage variants with SUD and MUD depending on the number of preambles S as a function of the arrival rate $\lambda$.}
	\label{fig:FigyAccess_latency}
	\vspace{-3pt}
\end{figure}
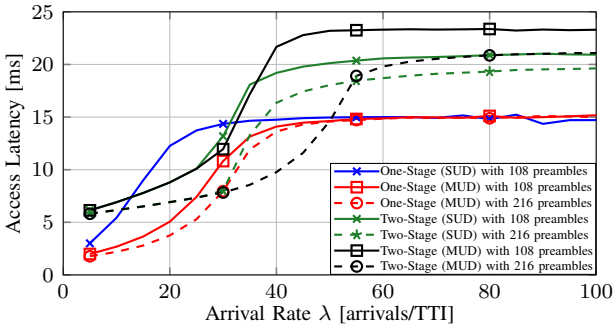

Figure~\ref{fig:FigyAccess_latency} shows the achievable access latency of successful packet transmissions with SUD and MUD. We see that the one-stage protocol overall can achieve significantly smaller delays if the traffic load is very low. A combination with MUD further reduces the access latency. The good result for very high load is misleading in this respect because the corresponding throughput in Fig.~\ref{fig:FigxProtocol_throughput} is close to zero. In the range around \tblue{$\lambda = 25$}, the two-stage protocol benefits from the lower collision probability, i.e.~smaller retransmission rate. We further see that a larger set of preambles $S$ can also provide some gain regarding access latency and that combination with MUD is advantageous as well.


\subsection{Signature based Access with Integrated Authentication \tblue{(SBAIA)}} 
\label{sub:signature_based_access_with_integrated_authentication}
\begin{figure}
\centering
	\includegraphics[width=\linewidth]{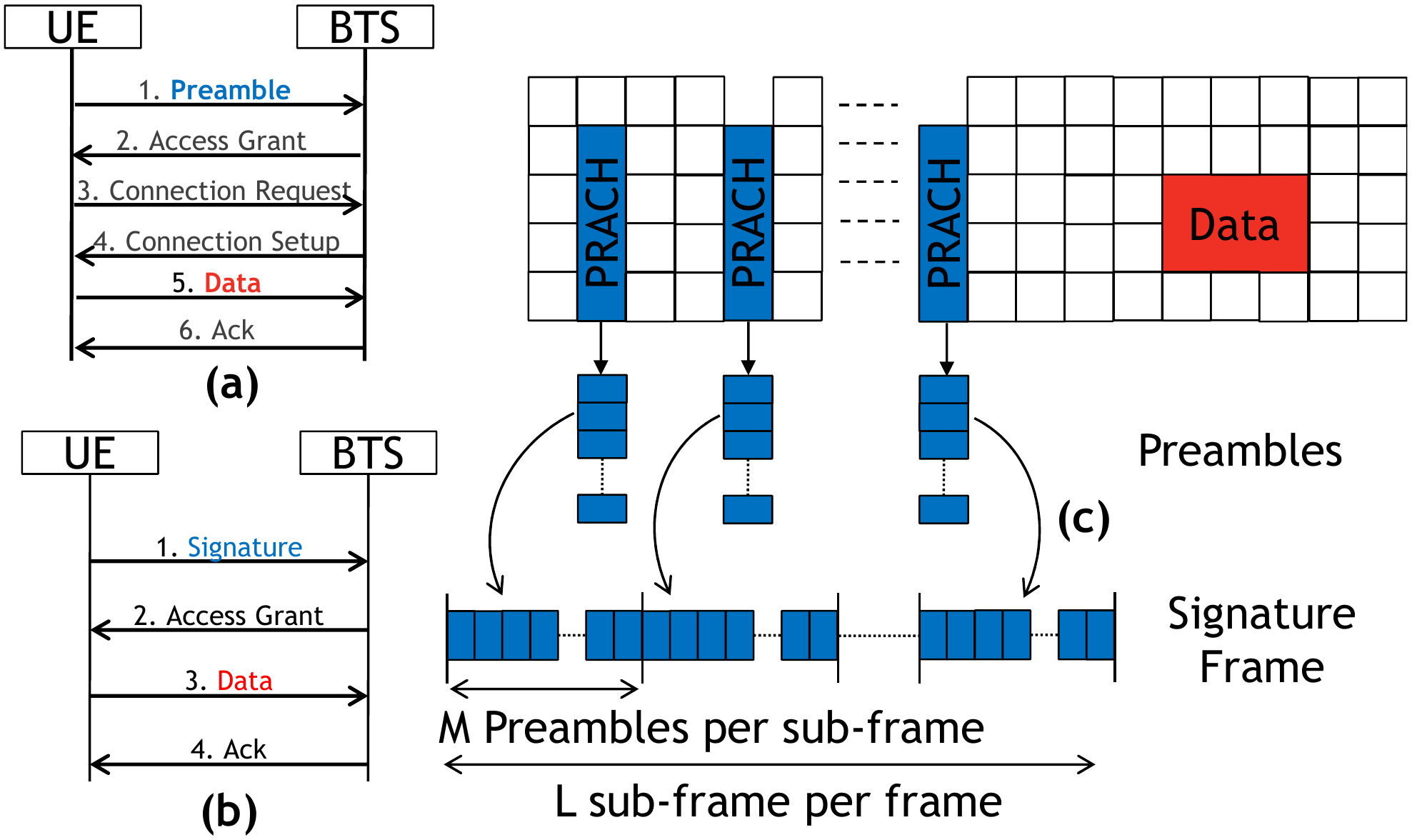}\caption{(a)
	LTE-based two-stages random access (b) Signature-based two-stages random
	access; (c) Physical random access resources mapping to random access
	preambles and Signature frame constructed from $L$ sub-frames composed each by
	$M$ random access preambles.}%
	\label{fig:SignatureBasedAccess}%
\end{figure}

In the LTE(-A/Pro) random access protocol, depicted in
Fig.~\ref{fig:SignatureBasedAccess}(a), each device contends for access within
a Physical Random Access Channel (PRACH) by selecting randomly one of the $M$
available preambles. In case the device's access attempt is not successful
(i.e.~the preamble selected by the device was also activated by at least one
other device or it was not detected at all), then the device will back-off and
re-attempt access later. This procedure is repeated until the device is either
successful or the amount of allowed retransmissions is exceeded. In case the
access attempt is successful, the device has then to inform the network about
its identity and how many resources it requires to transmit its data payload.
This protocol step is necessary only because the transmission of a preamble
does not encode any information about the device nor its requirements.

In contrast, in the proposed random access scheme, depicted in
Fig.~\ref{fig:SignatureBasedAccess}(b), we allow each device to contend with a
predefined sequence of preambles over multiple PRACHs, which we denote as the
device's \emph{signature}. These signatures, i.e.~the preamble activation
pattern over multiple PRACHs, are constructed based on information unique to
each device (such as the device's identity). From a protocol standpoint, this
signature can then be used to identify the device and its requirements (e.g.
the amount of resources required to transmit its data payload). This in turn
allows a significant reduction of the amount of exchanges in the access
protocol to achieve the same functionality, as it can be seen when comparing
Fig.~\ref{fig:SignatureBasedAccess}(a) and (b). These signatures are
transmitted synchronously over a frame composed of several PRACHs, as depicted
in Fig.~\ref{fig:SignatureBasedAccess}(c). This is made possible only if the
preambles in each PRACH: (i) are orthogonal to each other; (ii) can be
detected simultaneously; and (iii) allow the base station to detect a preamble
even when it is transmitted by multiple devices~\cite{ETT:ETT2656}, i.e.~a
collision in the ``preamble space'' is still interpreted as an activated
preamble. This last property can be interpreted as the OR logical operation,
since each preamble is detected as activated if there is \emph{at least} one
device that transmits the preamble. This observation was the motivation for
the use of Bloom filters -- a data structure based on the OR operation for
testing set membership~\cite{Bloom1970} -- for the construction of the access
signatures. Specifically, the device's identity is hashed over multiple
independent hash functions and the resulting output used to select which
preamble in which PRACH to activate. Finally, all the above properties can be
obtained from preambles generated from spread sequences such as the Zadoff-Chu sequences.

In the following we describe briefly the signature construction, transmission
and detection. Assume that a device's identity is given by $\mathbf{u}$ and
its corresponding signature as $\mathbf{s}^{(h)} = f(\mathbf{u})$. Where
$f(.)$ corresponds to the operation of hashing over multiple independent hash
functions. The resulting signature can be represented as a binary vector, in
such way that the bits at '0' correspond to inactive preambles, while bits at
'1' represent the active preambles. As the transmission of all the devices'
signatures occurs in a synchronous fashion, then the base station receiver
will observe a superposition of all the transmitted signatures as,
\begin{equation}
	\label{eq:y_new}\mathbf{y} = \bigoplus_{h = 1}^{N} \hat{\mathbf{s}}^{(h)},
\end{equation}
where $\hat{\mathbf{s}}^{(h)}$ is the detected version of $\mathbf{s}^{(h)}$.
The detection if a given signature is active is done by testing if the
following holds
\begin{align}\label{eq:det}
	\mathbf{s} = \mathbf{s} \bigotimes\mathbf{y},
\end{align}
where $\bigotimes$ is the bit-wise AND.

The drawback of this signature construction is that even in the case of
perfect preamble detection and no false detection, the base station can still
detect signatures that have not been transmitted (i.e.~the corresponding
device is not active) for which~\eqref{eq:det} holds. In other words, the base
station may decode \emph{false positives}. The signatures can then be designed
in terms of the number of active preambles and the signature length; and in
doing so control the number of false positives generated.

The signature decoding can be performed in an iterative manner, since the base
station will receive each PRACH sequentially; and compare each of the observed
active preambles with the valid signatures. This approach is inspired by the
fact that the active preambles, which constitute a signature, are randomly
spread over the PRACHs of the signature frame and, in principle, the base
station does not need to receive all of them to detect that the signature \tblue{has}
been transmitted. As the signature of a device is detected, the device is notified
and granted access to the channel, following the access protocol depicted in
Fig.~\ref{fig:SignatureBasedAccess}(a).

In the following we provide a comparison in terms of protocol throughput and access latency compared
with a LTE(-A/Pro) baseline. 
The PRACH configuration follows the details in Table~\ref{tab:params}.
The mean number of arrivals is assumed to be known, and the signature based scheme dimensioned for it. The probability of preamble detection by the base station is set to $p_{d}=0.99$ and the probability of
false detection of a preamble is set to $p_{f}=10^{-3}$ \cite{3GPPTS36.141}.
In the baseline, i.e.~LTE(-A/Pro) scheme, we assume the typical values for the backoff window of 20~ms, a maximum number of $10$ access attempts, 10~ms until the grant message is received and 40~ms until the connection setup (collision resolution) is received.
We assume that PRACH occurs every 1~ms, where there are $54$ available preambles for contention per PRACH in the LTE baseline which require 6 dedicated PRBs; while for the proposed scheme we assume that 216 preambles are available per PRACH that require 12 of the available PRBs for their generation.

\begin{figure}
\centering
%
%
%
%
\begin{tikzpicture}
\begin{axis}[%
small,
width=0.8\columnwidth,
height= 3.5cm,
scale only axis,
  try min ticks=6,
  max space between ticks=30pt,
  x tick label style={
    /pgf/number format/.cd,
    fixed,
    precision=2
  },
xmajorgrids,ymajorgrids,
xmin=0, xmax=3*38,
xlabel={Arrival Rate $\lambda$ [arrivals/TTI]},
ymin=0, ymax = 40,
ylabel={Protocol Throughput [users/TTI]},
legend cell align={left},
label style ={font=\footnotesize},
legend pos=south east,
legend style={font=\footnotesize}]

\addplot [mark=square,
mark repeat=5,
color=red,
solid,
x filter/.code={\pgfmathparse{\pgfmathresult*38}\pgfmathresult}]
coordinates{
(	0.1	,	4.80298005	)
(	0.2	,	9.6059601	)
(	0.3	,	14.40894015	)
(	0.4	,	19.2119202	)
(	0.5	,	24.01490025	)
(	0.6	,	28.8178803	)
(	0.7	,	33.62086035	)
(	0.8	,	38	)
(	0.9	,	38	)
(	1	,	38	)
(	1.1	,	38	)
(	1.2	,	38	)
(	1.3	,	38	)
(	1.4	,	38	)
(	1.5	,	38	)
(	1.6	,	38	)
(	1.7	,	38	)
(	1.8	,	38	)
(	1.9	,	38	)
(	2	,	38	)
(	2.1	,	38	)
(	2.2	,	38	)
(	2.3	,	38	)
(	2.4	,	38	)
(	2.5	,	38	)
(	2.6	,	38	)
(	2.7	,	38	)
(	2.8	,	38	)
(	2.9	,	38	)
(	3	,	38	)
};
\addlegendentry{Signature Lower Bound};
\addplot [mark=x,
mark repeat=5,
color=blue,
solid,
x filter/.code={\pgfmathparse{\pgfmathresult*38}\pgfmathresult}]
coordinates{
(	0.1	,	5	)
(	0.2	,	10	)
(	0.3	,	14.99945969	)
(	0.4	,	9.532800197	)
(	0.5	,	3.072930158	)
(	0.6	,	1.339639243	)
(	0.7	,	0.620757561	)
(	0.8	,	0.297827461	)
(	0.9	,	0.149382956	)
(	1	,	0.08419329	)
(	1.1	,	0.052016483	)
(	1.2	,	0.036845016	)
(	1.3	,	0.02729406	)
(	1.4	,	0.023159954	)
(	1.5	,	0.019432694	)
(	1.6	,	0.017206067	)
(	1.7	,	0.015727977	)
(	1.8	,	0.013922787	)
(	1.9	,	0.012419626	)
(	2	,	0.011002652	)
(	2.1	,	0.009901248	)
(	2.2	,	0.010162714	)
(	2.3	,	0.008981312	)
(	2.4	,	0.008527819	)
(	2.5	,	0.007542834	)
(	2.6	,	0.006942515	)
(	2.7	,	0.006539873	)
(	2.8	,	0.0057368	)
(	2.9	,	0.005606896	)
(	3	,	0.005053478	)

};
\addlegendentry{LTE Baseline};
\end{axis}
\end{tikzpicture}
	\caption{Protocol throughput for signature based access with 216 preambles.}
	\label{fig:AAU_ProtocolThroughput}
\end{figure}
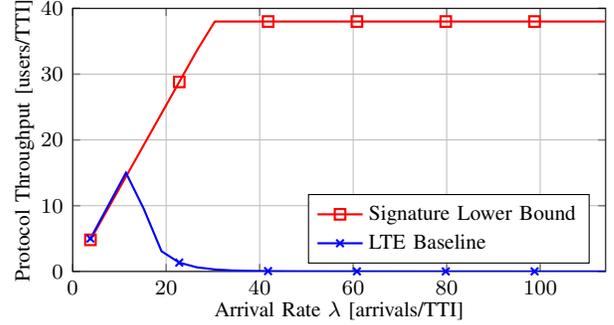

The protocol throughput achieved by this scheme is provided in Fig.~\ref{fig:AAU_ProtocolThroughput}. Note that the result provided is the lower bound throughput, yet for higher loads the throughput will not go beyond 38 packets per TTI as this corresponds to the maximum available PRBs per TTI.

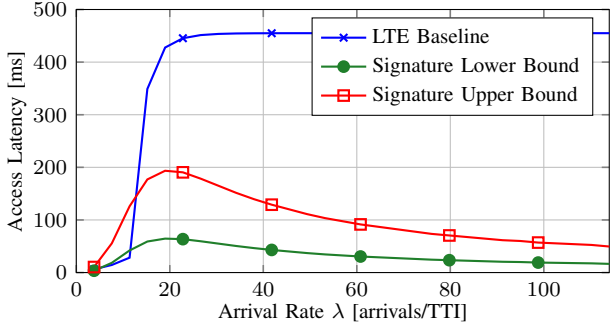
\begin{figure}
\centering
%
%
%
%

\begin{tikzpicture}
\begin{axis}[%
small,
width=0.8\columnwidth,
height= 3.5cm,
scale only axis,
  try min ticks=6,
  max space between ticks=30pt,
  x tick label style={
    /pgf/number format/.cd,
    fixed,
    precision=2
  },
xmajorgrids,ymajorgrids,
xmin=0, xmax=3*38,
xlabel={Arrival Rate $\lambda$ [arrivals/TTI]},
ymin=0, ymax = 500,
ylabel={Access Latency [ms]},
label style ={font=\footnotesize},
legend cell align={left},
legend pos= north east,
legend style={font=\footnotesize}]

\addplot [mark=x,
mark repeat=5,
color=blue,
solid,
x filter/.code={\pgfmathparse{\pgfmathresult*38}\pgfmathresult}]
coordinates{
(	0.1	,	6.271151435	)
(	0.2	,	14.42373157	)
(	0.3	,	28.27196832	)
(	0.4	,	348.9722347	)
(	0.5	,	427.6919158	)
(	0.6	,	445.4235809	)
(	0.7	,	451.3848186	)
(	0.8	,	453.5961569	)
(	0.9	,	454.4654274	)
(	1	,	454.7844295	)
(	1.1	,	454.9152614	)
(	1.2	,	454.9537775	)
(	1.3	,	454.9800007	)
(	1.4	,	454.9866005	)
(	1.5	,	454.9890216	)
(	1.6	,	455.0072995	)
(	1.7	,	454.9916488	)
(	1.8	,	454.9912378	)
(	1.9	,	454.9842477	)
(	2	,	454.9963698	)
(	2.1	,	455.0034546	)
(	2.2	,	454.9944133	)
(	2.3	,	455.0031569	)
(	2.4	,	454.9946333	)
(	2.5	,	455.0011983	)
(	2.6	,	454.9946548	)
(	2.7	,	454.9954436	)
(	2.8	,	455.0002322	)
(	2.9	,	454.9959049	)
(	3	,	455.0008485	)
};
\addlegendentry{LTE Baseline}

\addplot [
mark=*,
mark repeat=5,
color=darkgreen,
solid,
x filter/.code={\pgfmathparse{\pgfmathresult*38}\pgfmathresult}]
coordinates{
(	0.1	,	3.5	)
(	0.2	,	18.5	)
(	0.3	,	42	)
(	0.4	,	59	)
(	0.5	,	64.5	)
(	0.6	,	63.5	)
(	0.7	,	59.5	)
(	0.8	,	55	)
(	0.9	,	50.5	)
(	1	,	46.5	)
(	1.1	,	43	)
(	1.2	,	40	)
(	1.3	,	37	)
(	1.4	,	34.5	)
(	1.5	,	32.5	)
(	1.6	,	30.5	)
(	1.7	,	29	)
(	1.8	,	27.5	)
(	1.9	,	26	)
(	2	,	24.5	)
(	2.1	,	23.5	)
(	2.2	,	22.5	)
(	2.3	,	21.5	)
(	2.4	,	20.5	)
(	2.5	,	20	)
(	2.6	,	19	)
(	2.7	,	18.5	)
(	2.8	,	18	)
(	2.9	,	17.5	)
(	3	,	16.5	)

};
\addlegendentry{Signature Lower Bound}

\addplot [
mark=square,
mark repeat=5,
color=red,
solid,
x filter/.code={\pgfmathparse{\pgfmathresult*38}\pgfmathresult}]
coordinates{
(	0.1	,	10.5	)
(	0.2	,	55.5	)
(	0.3	,	126	)
(	0.4	,	177	)
(	0.5	,	193.5	)
(	0.6	,	190.5	)
(	0.7	,	178.5	)
(	0.8	,	165	)
(	0.9	,	151.5	)
(	1	,	139.5	)
(	1.1	,	129	)
(	1.2	,	120	)
(	1.3	,	111	)
(	1.4	,	103.5	)
(	1.5	,	97.5	)
(	1.6	,	91.5	)
(	1.7	,	87	)
(	1.8	,	82.5	)
(	1.9	,	78	)
(	2	,	73.5	)
(	2.1	,	70.5	)
(	2.2	,	67.5	)
(	2.3	,	64.5	)
(	2.4	,	61.5	)
(	2.5	,	60	)
(	2.6	,	57	)
(	2.7	,	55.5	)
(	2.8	,	54	)
(	2.9	,	52.5	)
(	3	,	49.5	)
};
\addlegendentry{Signature Upper Bound}

\end{axis}
\end{tikzpicture}
	\caption{Access latency of signature based access for 216 preambles.}
	\label{fig:AAU_AccessLatency}
\end{figure}

In Fig.~\ref{fig:AAU_AccessLatency} is provided the upper and lower bounds of the access latency achieved by the signature scheme, where it can be observed that both bounds decrease with the increasing arrival rate. This decrease is due to the signature length decreasing with the access load, which has a direct impact on the access latency.

Signature based random access is a novel access scheme that allows the
reduction of the exchanges required to transmit small payloads in wireless
access protocols. The functionality of the described protocol can be extended
to include authentication and security establishment and prioritization of
traffic~\cite{61f883fe4ffe46d88621c793a9289775,4d13fe2667c5453dab8c5ca158f12ef8}. This is possible, since the access pattern can be made in such a way to encode any kind of information.


\subsection{Non-Orthogonal Access with Time Alignment Free Transmission \tblue{(NOTAFT)}} 
\label{sub:non_orthogonal_access_with_time_alignment_free_transmission}

\tblue{In the current LTE system, both CP-OFDM and DFTs-OFDM impose strict synchronization requirements to the system. In order to guarantee reliable link performance, the timing inaccuracy of the receiving window needs to be kept within the range of the cyclic prefix. In the cellular uplink, however, the mobility of the users yields a continuous change in the propagation delay of their transmission signals, and thus introduces time-variant timing offsets. In order to tackle such random and variable timing misalignment, a closed-loop time alignment (TA) procedure is implemented in the LTE systems for enabling the BS to track each individual user's uplink timing during an active connection. However, for MTC with stringent power consumption limitations and sporadic activity with rather short data packets, it is desirable to design a simplified access procedure that can enable a grant-free and TA-free transmission of a short data packet in a single shot, yielding the one-stage access according to Fig.~\ref{fig:AccessProtocolClassification}.}

The first requirement derived from the above problem statement is the time asynchronous transmission, which is a feature \tblue{supported} by enhanced multi-carrier schemes like pulse shaped OFDM \cite{Zhao2016b}. Pulse-shaped OFDM (P-OFDM) fully maintains the signal structure of CP-OFDM, while allowing for pulse shapes other than the rectangular pulse to balance the localization of the signal power in the time and frequency domain. \tblue{Let $M$ be the FFT size, $N$ be the number of samples within one symbol period and $T_s$ be the sampling period.} We consider the time-frequency rectangular lattice for the OFDM system $(T, F)$, with $T=N T_s$ denoting the symbol period\tblue{ and $F=(M T_s)^{-1}$ the subcarrier spacing}. The P-OFDM transmit signal can be given as
\begin{equation}
	s(t)= \sum_{n=-\infty}^{+ \infty} \sum_{m=1}^{M} a_{m,n} g(t-n T) e^{j 2\pi m F(t-nT)}.
\end{equation}
Here, $a_{m,n}$ is the complex-valued information bearing symbol with sub-carrier index $m$ and symbol index $n$, respectively, and $g(t)$ represents the transmit pulse shape. At the receiver, demodulation of the received signal $r(t)$ is performed based on the receive pulse shape $\gamma(t)$: 
\begin{equation}
	\hat a_{m,n} = \int_{n=-\infty}^{+ \infty} r(t) \gamma(t-n T) e^{-j2\pi m F(t-nT)}.
\end{equation}
By carefully designing the pulse shapes $g(t)$ and $\gamma(t)$, the power localization in the time and frequency domain of a pulse can be adjusted. \tblue{In this work, robustness against distortions from large timing offsets is desired. To this end, following the design approach elaborated in \cite{Zhao2016a}, an orthogonalized Gaussian pulse which spreads four symbol periods is adopted as the transmit and receive pulse}. In comparison to CP-OFDM, it can be shown that this pulse exhibits \tblue{a high resilience against timing offsets}. This allows \tblue{for} asynchronous transmission without timing adjustment within cell coverage. Therefore, the timing alignment procedure during the random access phase can be omitted.

\tblue{In contrast to the baseline assumptions outlined in Table~\ref{tab:params}, we assume pulse shaped OFDM (P-OFDM) \cite{Zhao2016b} coupled with a space division multiple access (SDMA) scheme relying on multiple antennas at the BS. Coupling these two technologies facilitates a non-orthogonal grant-free access scheme supporting collision resolution based on MIMO detection techniques on the BS side. To this end, we assume that each spatial layer carries a demodulation reference signal (DMRS) orthogonal to those of the other layers.}

\begin{figure}
	\centering
	\includegraphics[width=\columnwidth]{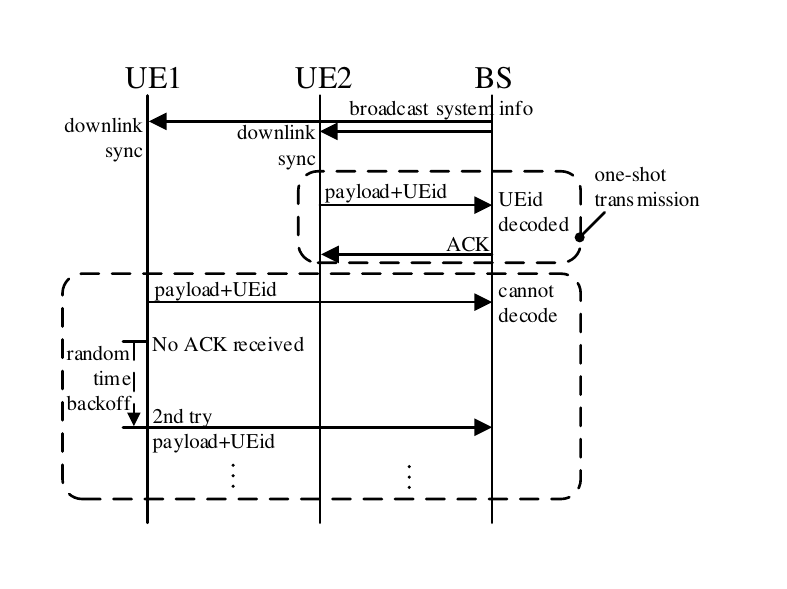}
	\caption{Proposed random access procedure with non-orthogonal time alignment free transmission}
	\label{fig:gfacess_proc}
\end{figure}

\tblue{The proposed random access scheme with non-orthogonal TA-free transmission is illustrated in Fig. \ref{fig:gfacess_proc} and can be described as follows:
\begin{enumerate}
	\item UE establishes downlink synchronization to the primary cell and obtains the system configuration by decoding broadcast channel information. The broadcast information may include cell-specific reference signal setting, maximum number of retransmission and default transmission scheme.
	\item The UE randomly selects a resource block and transmits its short packet data payload including its UE identifier. Here, the resource block consists of the time-frequency resource on a spatial layer which is identified by its DMRS.
	\item The BS decodes the received signal. With a successfully decoded data payload, the UE can be identified and an acknowledgment is fed back. 
	\item If a UE receives an ACK, the NOTAFT transmission is completed.
	\item If no ACK is received, a UE takes a random time backoff, and then steps 2-3 are repeated until either an ACK is received or the maximum number of retransmissions is reached.
\end{enumerate}}

\tblue{We examine the uplink transmission in a single macro-cell scenario without timing adjustment. Due to the radio propagation delay, a timing misalignment is present upon the arrival of the uplink signal at the BS. Assuming a cell radius of 2~km, this timing misalignment is calculated according to the propagation delay of the round trip, laying approximately in the range of $[0, 13] \, \mu s$. Link performance evaluation in \cite{Zhao2016b} shows no significant loss for such scenario when P-OFDM with an appropriately designed pulse spanning four symbol durations is employed. Therefore, for the protocol evaluation, we assume that packet loss is not caused by the timing misalignment, but only by the resource collision, i.e.~if two UEs select the same spatial layer on the same resource block. Since access preamble is used, the total number of available resource blocks, i.e.~$50$ PRBs, can be employed for non-orthogonal data transmission for 10~MHz mode. With a typical setting of four antennas on the BS side, this amounts to a total of $200$ random access opportunities per TTI. This scheme is compared to the multi-stage access scheme with TA, depicted in Fig.~\ref{fig:AccessProtocolClassification}. Parameters listed in Table~\ref{tab:params} are applied.}

\begin{figure}
	\centering
%
%
%
%
\begin{tikzpicture}
\begin{axis}[%
small,
width=0.8\columnwidth,
height= 3.5cm,
scale only axis,
  try min ticks=6,
  max space between ticks=30pt,
  x tick label style={
    /pgf/number format/.cd,
    fixed,
    precision=2
  },
xmajorgrids,ymajorgrids,
xmin=0, xmax=2*50,
xlabel={Arrival Rate $\lambda$ [arrivals/TTI]},
ymin=0, ymax = 100,
ylabel={Protocol Throughput [users/TTI]},
label style ={font=\footnotesize},
legend pos=north west,
legend style={font=\footnotesize}]

\addplot [mark=x,
color=blue,
solid,
x filter/.code={\pgfmathparse{\pgfmathresult*50}\pgfmathresult}]
coordinates{
(	0.02	,	1	)
(	0.10	,	4.9979	)
(	0.20	,	9.9494	)
(	0.40	,	19.0451	)
(	1.00	,	30.1363	)
(	2.00	,	11.1572	)
};
\addlegendentry{Multi-stage access};

\addplot [mark=square,
color=red,
solid,
x filter/.code={\pgfmathparse{\pgfmathresult*50}\pgfmathresult}]
coordinates{
(	0.02	,	1	)
(	0.10	,	5	)
(	0.20	,	10	)
(	0.40	,	20	)
(	1.00	,	49.4354	)
(	2.00	,	90.1927	)
};
\addlegendentry{Proposed RA};
\end{axis}
\end{tikzpicture}
	\caption{Protocol throughput of the non-orthogonal access with time alignment free transmission}
	\label{fig:thruput}
\end{figure}
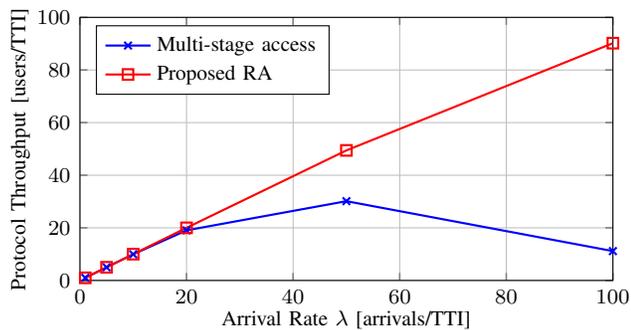

\tblue{Fig.~\ref{fig:thruput} depicts the achievable packet throughput as a function of the arrival rate. Since no resource is allocated for the random access procedure, all PRBs are utilized for data transmission. Given a much higher number of random access opportunities, the proposed NOTAFT scheme offers significantly higher throughput especially when the arrival rate is relatively high.}

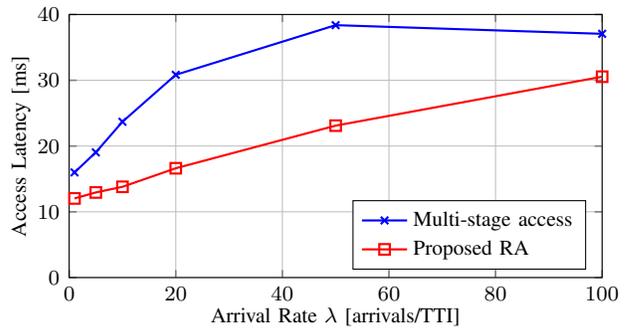
\begin{figure}
	\centering
%
%
%
%

\begin{tikzpicture}
\definecolor{darkgreen}{rgb}{0.12549019607843137255,0.4980392156862745098,0.16862745098039215686}
\begin{axis}[%
small,
width=0.8\columnwidth,
height= 3.5cm,
scale only axis,
  try min ticks=6,
  max space between ticks=30pt,
  x tick label style={
    /pgf/number format/.cd,
    fixed,
    precision=2
  },
xmajorgrids,ymajorgrids,
xmin=0, xmax=2*50,
xlabel={Arrival Rate $\lambda$ [arrivals/TTI]},
ymin=0, ymax =40,
ylabel={Access Latency [ms]},
label style ={font=\footnotesize},
legend pos=south east,
legend style={font=\footnotesize}]

\addplot [mark=x,
color=blue,
solid,
x filter/.code={\pgfmathparse{\pgfmathresult*50}\pgfmathresult}]
coordinates{
(	0.02	,	15.99	)
(	0.10	,	19.04	)
(	0.20	,	23.69	)
(	0.40	,	30.83	)
(	1.00	,	38.38	)
(	2.00	,	37.06	)
};
\addlegendentry{Multi-stage access};

\addplot [
mark=square,
color=red,
solid,
x filter/.code={\pgfmathparse{\pgfmathresult*50}\pgfmathresult}]
coordinates{
(	0.02	,	12.03	)
(	0.10	,	12.94	)
(	0.20	,	13.8	)
(	0.40	,	16.62	)
(	1.00	,	23.1	)
(	2.00	,	30.54	)
};
\addlegendentry{Proposed RA};

\end{axis}
\end{tikzpicture}
	\caption{Access latency of the non-orthogonal access with time alignment free transmission}
	\label{fig:latency}
\end{figure}

As shown in Fig.~\ref{fig:latency}, since the timing adjustment procedure is removed, the proposed one-shot transmission scheme exhibits lower access latency compared to the baseline approach. 

In summary, the proposed access procedure facilitates a 'single-shot transmission’, enabling a reduced end-to-end latency as well as a lower signaling overhead for short packet transmissions. Thanks to this, it could substantially extend the battery life of \tblue{devices} for a better sleep/wake-up operation.	



\section{PHY and MAC Integrated Schemes} 
\label{sub:phy_and_mac_integrated_schemes}
\tblue{In the following we present four approaches that extend the pure MAC protocol view of the previous section in terms of the physical layer assumptions. Here, all presented results include simulation of physical layer transmission at least including coding and modulation and in most cases also channel estimation. First, we present Compressive Sensing Multi-User Detection (CSMUD) which exploits sparsity due to sporadic activitiy in mMTC enabling efficient Multi-User detection in each random access slot of a slotted ALOHA setting. Second, we present Coded Random Access with Physical Layer Network Coding (CRAPLNC) extending the CSMUD approach to frames using ideas from network coding, which results in a high throughput Coded Random Access scheme. Third, we present Compressive Sensing Coded Random Access (CCRA) that combines Coded Random Access CSMUD with an underlay control channel significantly reducing control overhead. Finally, we present Slotted Compute and Forward (SCF) focusing on very dense networks with high numbers of mini base stations forwarding messages to a full base station to efficiently enable mMTC scenarios.}

\subsection{Compressive Sensing Multi-User Detection \tblue{(CSMUD)}} 
\label{sub:compressive_sensing_multi_user_detection}
\tblue{The massive access problem outlined in section~\ref{sec:challenges} is characterized by a massive number of MTDs that do not send information continously but rather sporadically in large time intervals or even event driven. As already outlined from a MAC perspective different access protocols can structure such a sporadic access pattern. Still, the physical layer design of the access procedure remains open and naturally depends on the MAC protocol choice. Focusing on a one-stage protocol early works on sporadic access in combination with Code Division Multiple Access (CDMA) already noted that intermittend user activity leads to a multi-user detection (MUD) problem with sparsity that required novel algorithmic solutions~\cite{sparse_mu_detection}. Most importantly, with the development of compressive sensing (CS) a new mathematical tool was available to solve MUD with sparsity~\cite{candes2006compressivesampling}. A major advantage of combininig compressive sensing ideas and MUD lies in the theoretical guarantees of CS for under-determined detection problems. Prior to the so-called Compressive Sensing Multi-user Detection (CSMUD) most MUD problems with sparsity focued on fully-determined systems where the number of resources and users coincide. With CS detection guarantees can be given even if the number of resources is strictly smaller than the number of users which enables user detection even in highly overloaded CDMA setups. From this basic idea CSMUD has been extended in multiple directions ranging from non-coherent communication \cite{monsees2015a} to channel estimation with user activity detection. In the following we will revisit the CSMUD ideas for channel estimation with simultaneous user actvity detection \cite{schepker2013} and present numerical evaluation result in combination with a simple one-shot protocol. The basic CSMUD ideas presented in the following also serve as an introduction to the presented solutions in section~\ref{sub:coded_random_access_with_PLNC} and \ref{sub:compressive_sensing_coded_random_access}.}

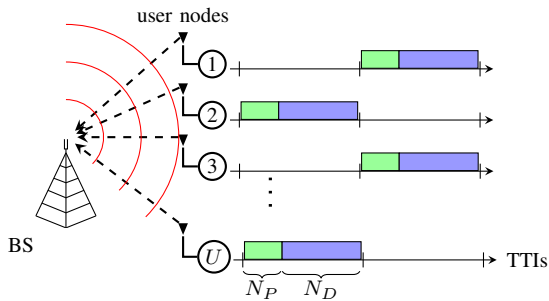
\begin{figure}[ptb]
	\centering
	\begin{tikzpicture}[font=\footnotesize]

\tikzset{
se/.style={-stealth, thick},
dots/.style={dash pattern=on 1pt off 4pt,very thick},
box/.style={rectangle, fill=white, inner sep=7pt, very thick, draw,drop shadow={fill=black, xshift=1mm, yshift=-1mm}},
hbox/.style={rectangle, color=white, fill=white, inner sep=7pt, very thick, draw,drop shadow={fill=white, xshift=1mm, yshift=-1mm}},
vbox/.style={box,rotate=90,text centered},
pbox/.style={rectangle, fill=green!40, minimum width=0.5cm, draw},
dbox/.style={rectangle, fill=blue!40, minimum width=1.05cm, draw},
antenna/.style={isosceles triangle,fill=black, shape border rotate=-90, inner sep=1pt},
sporadic/.style={-stealth,thick,dashed},
axis/.style={-stealth, thin},
axistick/.style={thin,minimum height=6mm,path picture={\draw[black,thin] (0,1mm) -- (0,-1mm);}},
data/.style={circle,thick,draw, inner sep=1.2pt},
basestation/.style={minimum height=1.2cm,minimum width=1cm,path picture={
\draw[black,thin] (0,-6mm) -- (0,4mm);
\draw[black,thin] (-4mm,-4.4mm) -- (0,4mm);
\draw[black,thin] (4mm,-4.4mm) -- (0,4mm);
\draw[black,thin] (4mm,-4.4mm) -- (0,-6mm) -- (-4mm,-4.4mm);
\draw[black,thin] (3.2mm,-2.7mm) -- (0,-4mm) -- (-3.2mm,-2.7mm);
\draw[black,thin] (2.4mm,-1.0mm) -- (0,-2mm) -- (-2.4mm,-1.0mm);
\draw[black,thin] (1.6mm,0.6mm) -- (0,0mm) -- (-1.6mm,0.6mm);
\draw[black,thin] (0.8mm,2.3mm) -- (0,2mm) -- (-0.8mm,2.3mm);
\draw[black] (0.2mm,5.8mm) -- (0.2mm,4.5mm) -- (0,4.5mm) -- (-0.2mm,4.5mm) -- (-0.2mm,5.8mm);
\draw[black,thin] (0,4.5mm) -- (0mm,4mm);
}},
}

 \node[basestation] (bs) at (0cm,0cm){};
 \node[below right=0cm and -1.4cm of bs, text width=5.8cm, text justified] (bt){BS};

 \node[antenna,right=1.5cm of bs.north] (node3) {};
 \coordinate[right=0.15cm of bs.north] (t3);
 \draw[sporadic] (node3) -- (t3);
 \node[data,below right=0.3cm of node3] (data3) {3};
 \draw[thick] (data3) -| (node3);
 \coordinate[above right=-0.2cm and 0.1cm of data3] (start3);
 \coordinate[right=3.5cm of start3] (stop3);
 \draw[axis] (start3) -- (stop3);
 \node[axistick, right= 0.0cm of start3] () {};
 \node[axistick, right= 1.6cm of start3] () {};
 \node[axistick, right= 3.2cm of start3] () {};
 \node[pbox, above right= 0.001cm and 1.74cm of start3] () {};
 \node[dbox, above right= 0.001cm and 2.24cm of start3] () {};

 \node[antenna,above=0.5cm of node3.north] (node2) {};
 \coordinate[above right= 0.045 and 0.135cm of bs.north] (t2);
 \draw[sporadic] (node2) -- (t2);
 \node[data,below right=0.3cm of node2] (data2) {2};
 \draw[thick] (data2) -| (node2);
 \coordinate[above right=-0.20cm and 0.1cm of data2] (start2);
 \coordinate[right=3.5cm of start2] (stop2);
 \draw[axis] (start2) -- (stop2);
\node[axistick, right= 0.0cm of start2] () {};
 \node[axistick, right= 1.6cm of start2] () {};
 \node[axistick, right= 3.2cm of start2] () {};
 \node[pbox, above right= 0.001cm and 0.14cm of start2] () {};
 \node[dbox, above right= 0.001cm and 0.64cm of start2] () {};

 \node[antenna,above=0.5cm of node2.north] (node1) {};
 \coordinate[above right=0.085cm and 0.12cm of bs.north] (t1);
 \draw[sporadic] (node1) -- (t1);
 \node[data,below right=0.3cm of node1] (data1) {1};
 \draw[thick] (data1) -| (node1.south);
 \coordinate[above right=-0.2cm and 0.1cm of data1] (start1);
 \coordinate[right=3.5cm of start1] (stop1);
 \draw[axis] (start1) -- (stop1);
\node[axistick, right= 0.0cm of start1] () {};
 \node[axistick, right= 1.6cm of start1] () {};
 \node[axistick, right= 3.2cm of start1] () {};
 \node[pbox, above right= 0.001cm and 1.74cm of start1] () {};
 \node[dbox, above right= 0.001cm and 2.24cm of start1] () {};
 \node[above=0cm of node1] (sn){user nodes};

 \node[antenna,below=1.0cm of node3.south] (node4) {};
 \coordinate[below right=0.085cm and 0.12cm of bs.north] (t4);
 \draw[sporadic] (node4) -- (t4);
 \node[data,below right=0.3cm of node4] (data4) {$U$};
 \draw[thick] (data4) -| (node4);
 \coordinate[above right=-0.2cm and 0.1cm of data4] (start4);
 \coordinate[right=3.5cm of start4] (stop4);
 \draw[axis] (start4) -- (stop4);
 \node[right=0.1mm of stop4] () {\footnotesize TTIs};
 \node[axistick, right= 0.0cm of start4] () {};
 \node[axistick, right= 1.6cm of start4] () {};
 \node[axistick, right= 3.2cm of start4] () {};
 \node[pbox, above right= 0.001cm and 0.14cm of start4] () {};
 \node[dbox, above right= 0.001cm and 0.64cm of start4] () {};

 \coordinate[below right=0.4cm and 1.16cm of node3.south] (dots_start);
 \coordinate[above right=0.1cm and 1.16cm of node4.north] (dots_end);
 \draw[dots] (dots_start) -- (dots_end);

\draw[red] (bs.north) ++(-45:5mm) arc (-45:90:5mm);
\draw[red] (bs.north) ++(-45:10mm) arc (-45:90:10mm);
\draw[red] (bs.north) ++(-45:15mm) arc (-45:90:15mm);

\draw[decorate,decoration=brace] let \p1=(start4) in (\x1+0.63cm,\y1-1.5mm) -- node[midway,below]{$N_P$} (\x1+0.13cm,\y1-1.5mm) ;

\draw[decorate,decoration=brace] let \p1=(start4) in (\x1+1.68cm,\y1-1.5mm) -- node[midway,below]{$N_D$} (\x1+0.65cm,\y1-1.5mm) ;
\end{tikzpicture}
 \caption{Sporadic uplink transmission of multiple devices sending $N_P$ pilot sybmols and $N_D$ data symbols to a BS.}
	\label{fig:csmudsys}
\end{figure}

Following the assumptions laid out earlier, i.e.~a certain time and frequency budget is allocated to the MMC service and it is well separated and robust by choice of an appropriate waveform, Fig.~\ref{fig:csmudsys} depicts a schematic view on the MMC access protocol. \tblue{Each TTI all $N_{\text{act}}$ active users out of the overall $U$ users access the system by transmitting $N_P$ pilots and $N_D$ data symbols both spread over the whole bandwidth through one of $N_S$ pseudo-noise (PN) spreading sequences $\mathbf{s}_i\in\mathbb{C}^{L_S}\;\forall\;i=1,\dotsc,N_S$. The number of available spreading sequences $N_S$ and their length $L_S$ determine the physical layer performance of CSMUD. If the number of active users $N_{\text{act}}$ is in the order of or larger than the number of available spreading sequence $N_S$ collisions will occurr. If the less active users access the system than spreading sequences are available the systems performance will be dominated by the CSMUD performance, i.e.~the separation of the $N_S$ PN sequences of length $L_S$. Obviously, the longer the spreading sequence, the lower the achievable data rate given TTI length and bandwidth from Table~\ref{tab:params} but the higher the robustness and separability of spreading sequences. The resulting trade-off between MUD performance and collision probabilty in dependence of retransmission is highly non-trivial. Only the physical layer design trade-off between $N_P$ and $N_D$ was already investigated \cite{schepker2013}, but the interaction with different MAC protocols is still an open problem. Hence, we will restrict the presented evaluation results to a single parametrization that is designed to achieve the packet size of Table~\ref{tab:params}.}

\tblue{As indicated in Fig.~\ref{fig:csmudsys} each user sends a packet of two parts. The first part consists of $N_P$ pilot symbols that are unique per user and serve to estimate channel and activity through CSMUD. The second part consists of the spread $N_D$ data symbols that can be detected and decoded through standard approaches. Each slot is assumed to occupy 10~MHz and 1~ms per table~\ref{tab:params}. To formalize the task of CSMUD we summarize the user channels $\mathbf{h}_i\in\mathcal{C}^{N_h}\;\forall i=1,\dotsc,U$ in a stacked channel vector $\mathbf{h}=[\mathbf{h}_1,\dotsc,\mathbf{h}_U]^T \in\mathcal{C}^{N N_h}$. Due to the sporadic activity, channels of inactive users will be modeled as zeros, i.e.~for inactive user $\mathbf{h}_i=\mathbf{0}_{N_h} \;\forall i\in\bar{\mathbb{Z}}$, where $\bar{\mathbb{Z}}$ and $\mathbb{Z}$ denote the index set of all active and inactive users, respectively. This leads to additional structure in the detection problem, i.e.~the vector $\mathbf{h}$ is strictly group-sparse with groups of size $N_h$. The joint channel and activity signal model is then,}

\begin{equation}
	\mathbf{y} = \mathbf{S}\mathbf{h} + \mathbf{n},
\end{equation}

\tblue{where $\mathbf{S}\in\mathcal{C}^{M\times N}$ denotes the preamble matrix containing all user preambles, $\mathbf{y}\in\mathcal{C}^M$ denotes the received signal consisting of the superimposed $N_P$ pilots of all users at the base station and $\mathbf{n}\in\mathcal{C}^M$ summarize all noise sources as AWGN. The preamble matrix $\mathbf{S}$ exhibits a Toeplitz structure per user describing the convolution of channel and pilots, i.e.~$\mathbf{S}=[\mathbf{S}_i,\dotsc \mathbf{S}_U]^T$ with $\mathbf{S}_i$ being a Toeplitz matrix of user $i$ pilots $\mathbf{s}_i$.}

Depending on the underlying system assumptions (asynchronicity, waveform, channel model, etc.) \tblue{the exact values of $M$ and $N$ vary, but are dependent on the number of users $U$, the pilot length $N_P$ and the lenght of the channel impulse response $N_h$. To simplify notation we focus on a one-tap Rayleigh fading channel, i.e.~$N_h=1$. Then, the detection problem can be cast as}

\begin{equation} \label{eq:csdet}
	\hat{\mathbf{h}} = \argmin_{\mathbf{h}\in\mathcal{C}^{N}%
}\left\|  \mathbf{h}\right\|  _{0}\quad\text{s.t.}\quad\left\|  \mathbf{y}%
-\mathbf{Sh}\right\|  _{2}<\epsilon,
\end{equation}

which is easily extended to $N_h$-tap Rayleigh fading channels if a group sparsity constraint is introduced (cf.~\cite{schepker2013}).
The minimization in \eqref{eq:csdet} targets the sparsest vector denoted by the \lq\lq
{pseudo-Norm}\rq\rq$\left\|\mathbf{h}\right\|_{0}$ that counts the number of non-zeros given an $\ell_{2}$-norm constraint to adhere to a given noise level dependent on $\epsilon$. The solution of \eqref{eq:csdet} can be approached in many different ways like
convex relaxation or sub-optimal Greedy approaches which are meanwhile very well covered in the literature. 

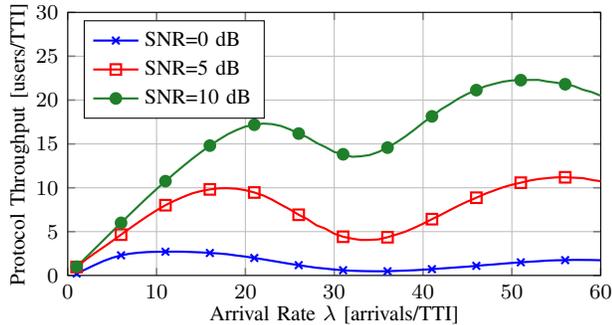
\begin{figure}[ptb]
	\centering
%
%
%
%
\begin{tikzpicture}
\begin{axis}[%
small,
width=0.8\columnwidth,
height= 3.5cm,
scale only axis,
  try min ticks=6,
  max space between ticks=30pt,
  x tick label style={
    /pgf/number format/.cd,
    fixed,
    precision=2
  },
xmajorgrids,ymajorgrids,
xmin=0, xmax=60,
xlabel={Arrival Rate $\lambda$ [arrivals/TTI]},
ymin=0, ymax = 30,
ylabel={Protocol Throughput [users/TTI]},
legend cell align={left},
label style ={font=\footnotesize},
legend pos=north west,
legend style={font=\footnotesize}]

\addplot [mark=x,
mark repeat=5,
color=blue,
solid,
x filter/.code={\pgfmathparse{\pgfmathresult}\pgfmathresult}]
coordinates{
 (1,0.22328231157639)(2,0.691476126212936)(3,1.17900612906504)(4,1.63313729833041)(5,2.00755392673444)(6,2.29245732275625)(7,2.48179847845491)(8,2.59659808231237)(9,2.66703631291425)(10,2.69602887960566)(11,2.70678919055155)(12,2.71556596647914)(13,2.70724854865808)(14,2.66880802486463)(15,2.62801104456313)(16,2.56194648603476)(17,2.50616074338997)(18,2.41500790144127)(19,2.2715391868253)(20,2.14936960825454)(21,1.99214998337243)(22,1.83730197207014)(23,1.66694470949192)(24,1.49550600512047)(25,1.32501673317701)(26,1.17353701283302)(27,1.01978910193751)(28,0.876742763924231)(29,0.782834334044931)(30,0.67306274534851)(31,0.594349485177849)(32,0.53554130541312)(33,0.500359815102097)(34,0.48299734944364)(35,0.475546330559338)(36,0.48080156101022)(37,0.522648313535553)(38,0.553759487780141)(39,0.592832201832156)(40,0.656495113342226)(41,0.716308941457785)(42,0.789792869740157)(43,0.860445935456858)(44,0.933458303421677)(45,1.00333013237387)(46,1.09770292674421)(47,1.18495361036242)(48,1.25686430678102)(49,1.34519835188674)(50,1.42356838616747)(51,1.49971896396709)(52,1.56227715123845)(53,1.62207895180997)(54,1.69141745193287)(55,1.72898929663579)(56,1.74624799340151)(57,1.77312868698453)(58,1.76793718327514)(59,1.75923660139558)(60,1.73469520705647) 
};
\addlegendentry{SNR=0~dB};
\addplot [mark=square,
mark repeat=5,
color=red,
solid,
x filter/.code={\pgfmathparse{\pgfmathresult}\pgfmathresult}]
coordinates{
 (1,0.96060010378461)(2,1.77999342070189)(3,2.51233073256235)(4,3.20190894307257)(5,3.91609128111595)(6,4.65253771242896)(7,5.38143841336335)(8,6.09309048054364)(9,6.77985010618995)(10,7.42534496255689)(11,8.01100569705936)(12,8.50709789450666)(13,8.95957684587114)(14,9.35634527616399)(15,9.6237830146011)(16,9.8123923142749)(17,9.93507285600838)(18,9.951565631329)(19,9.89125131156461)(20,9.72388446655626)(21,9.46116353195638)(22,9.12419277286586)(23,8.64476920088426)(24,8.0763964231438)(25,7.53657603995131)(26,6.91939508792516)(27,6.40330904571531)(28,5.71878270567498)(29,5.33414702872563)(30,4.75834811613012)(31,4.41383645466395)(32,4.20061613702715)(33,4.0423778957387)(34,4.05153740355351)(35,4.11162253841011)(36,4.35628244155534)(37,4.66756228181209)(38,4.99017970731922)(39,5.42696997353579)(40,5.89738526495344)(41,6.41822701877976)(42,6.91085327895455)(43,7.44682115685828)(44,7.95944060269557)(45,8.43312638608427)(46,8.86449444812777)(47,9.31621935721249)(48,9.68822698181503)(49,10.0553364489728)(50,10.3441957378037)(51,10.5771598756659)(52,10.8235076748168)(53,10.9955302690899)(54,11.1018415142189)(55,11.1870378790395)(56,11.1933542249312)(57,11.1449467466598)(58,11.0999182310617)(59,10.889790643922)(60,10.7465963807486) 
};
\addlegendentry{SNR=5~dB};
\addplot [mark=*,
mark repeat=5,
color=darkgreen,
solid,
x filter/.code={\pgfmathparse{\pgfmathresult}\pgfmathresult}]
coordinates{
 (1,1)(2,2)(3,3)(4,3.99986443958133)(5,4.99924376588165)(6,5.9957965887927)(7,6.98586066339132)(8,7.96532418815978)(9,8.92202361005037)(10,9.85052597612075)(11,10.7532677797349)(12,11.6252498106101)(13,12.4733377412907)(14,13.2929250601144)(15,14.0584584196989)(16,14.8069679653347)(17,15.4780485451718)(18,16.0931602591253)(19,16.5789707200529)(20,16.9781959612512)(21,17.1869972171029)(22,17.3214489666463)(23,17.2588504904495)(24,17.0468609746661)(25,16.6619448186839)(26,16.1796225960574)(27,15.741010574414)(28,15.1212245631616)(29,14.6446697740312)(30,14.0857934086209)(31,13.8163481956458)(32,13.5331592527007)(33,13.5890789907189)(34,13.7356231371816)(35,14.0950097597828)(36,14.5764697815216)(37,15.1789102427255)(38,15.8487471438797)(39,16.6049281813722)(40,17.4126055530355)(41,18.1412774067176)(42,18.9270762314458)(43,19.5288258771036)(44,20.1780145945401)(45,20.7056396657009)(46,21.1349482309044)(47,21.5719039096122)(48,21.8461704097302)(49,22.0999879203246)(50,22.2344513037543)(51,22.2714109672091)(52,22.2947216316293)(53,22.2985949066616)(54,22.1115460368315)(55,21.9857148313269)(56,21.8052512942709)(57,21.5859356124614)(58,21.1812380472493)(59,20.9079317535692)(60,20.4971362443965) 
};
\addlegendentry{SNR=10~dB};
\end{axis}
\end{tikzpicture}\caption{Protocol throughput for CSMUD.}
	\label{fig:csmudPT}
\end{figure}

\begin{figure}[ptb]
	\centering
%
%
%
%

\begin{tikzpicture}
\begin{axis}[%
small,
width=0.8\columnwidth,
height= 3.5cm,
scale only axis,
  try min ticks=6,
  max space between ticks=30pt,
  x tick label style={
    /pgf/number format/.cd,
    fixed,
    precision=2
  },
xmajorgrids,ymajorgrids,
xmin=0, xmax=60,
xlabel={Arrival Rate $\lambda$ [arrivals/TTI]},
ymin=0, ymax = 25,
ylabel={Access Latency [ms]},
label style ={font=\footnotesize},
legend pos= north west,
legend style={font=\footnotesize}]

\addplot [mark=x,
mark repeat=5,
color=blue,
solid,
x filter/.code={\pgfmathparse{\pgfmathresult}\pgfmathresult}]
coordinates{
 (1,12.3242694682247)(2,11.1159868099622)(3,10.9188142733922)(4,11.1183636130481)(5,11.5344408026655)(6,11.9757971473393)(7,12.4190977651149)(8,12.872630707654)(9,13.1696479670882)(10,13.4402510855262)(11,13.6838404196824)(12,13.8999376286491)(13,14.0629388744857)(14,14.2219613467883)(15,14.4406312893917)(16,14.5468002516131)(17,14.6554214086026)(18,14.9219993038067)(19,14.932888227939)(20,15.1537673317278)(21,15.2288193875946)(22,15.4079984098529)(23,15.6455880710598)(24,15.6763905762023)(25,15.7111198619448)(26,15.8378758331437)(27,15.7937355833707)(28,15.8623292167321)(29,16.1369339033297)(30,15.9443144339551)(31,15.9362504601769)(32,16.0954428782111)(33,16.0077987104636)(34,16.135599645369)(35,15.9614842496896)(36,16.0063803290742)(37,16.1872777836514)(38,15.9337961903756)(39,16.1682079386334)(40,15.9594877609649)(41,16.1885394043655)(42,16.0878767995309)(43,15.9843657347787)(44,16.1101344557564)(45,16.0542175846046)(46,15.977382529473)(47,16.0935281746935)(48,15.9804671457739)(49,15.9630663266934)(50,15.9434499445736)(51,15.941301920823)(52,15.8321924981149)(53,15.9181402158925)(54,15.9143573389639)(55,15.8193113527021)(56,15.8566618323843)(57,15.9303137556276)(58,15.9032373251544)(59,15.9166811008135)(60,15.9698024309767) 
};
\addlegendentry{SNR=0~dB};
\addplot [
mark=square,
mark repeat=5,
color=red,
solid,
x filter/.code={\pgfmathparse{\pgfmathresult}\pgfmathresult}]
coordinates{
 (1,4.24801965275395)(2,5.16821584234119)(3,5.6014720969929)(4,5.82315044025118)(5,5.89769236947117)(6,5.93757215029582)(7,6.00289524019878)(8,6.1132054627304)(9,6.25962775061082)(10,6.42877270435864)(11,6.66665926279911)(12,6.96585216540618)(13,7.26489325874613)(14,7.59532598635765)(15,8.02275260502474)(16,8.42646575717986)(17,8.86467426323057)(18,9.39442637259036)(19,9.83989173308228)(20,10.3725959275423)(21,10.8878515720686)(22,11.4541652931958)(23,11.9148728229628)(24,12.4761183652001)(25,12.9690673287911)(26,13.3930141451218)(27,13.7639702972047)(28,14.1210566272414)(29,14.3319645290067)(30,14.5942627000204)(31,14.7915522332117)(32,14.911241260816)(33,14.9352147058913)(34,14.8601714792822)(35,14.9547781676966)(36,14.8745915970435)(37,14.7436125648441)(38,14.7104694641111)(39,14.5519105317034)(40,14.5101411537966)(41,14.3783585513524)(42,14.2569429636276)(43,14.1412276901625)(44,14.0576476378377)(45,13.9971673879051)(46,13.954147188431)(47,13.8611121864334)(48,13.8712799417295)(49,13.8127805684312)(50,13.8095555217198)(51,13.8018519903425)(52,13.8028834242448)(53,13.8085083535371)(54,13.8610068760669)(55,13.8795158884322)(56,13.9358398409742)(57,14.0253351134292)(58,14.0784461007421)(59,14.1948540643266)(60,14.2659545468359) 
};
\addlegendentry{SNR=5~dB};
\addplot [
mark=*,
mark repeat=5,
color=darkgreen,
solid,
x filter/.code={\pgfmathparse{\pgfmathresult}\pgfmathresult}]
coordinates{
 (1,3.50211714513998)(2,3.50106607500168)(3,3.50090470703117)(4,3.50060179038001)(5,3.50191678790179)(6,3.51468276271145)(7,3.53136153677201)(8,3.56516566325777)(9,3.61435822267805)(10,3.68580800598747)(11,3.7613231255948)(12,3.84110519752296)(13,3.93785030280311)(14,4.0356218526492)(15,4.16561358425376)(16,4.2965805146809)(17,4.4713024048097)(18,4.67407604030254)(19,4.94446383548953)(20,5.26789321357263)(21,5.67553322859405)(22,6.11644352299137)(23,6.64812139290875)(24,7.22847113424234)(25,7.87879047248393)(26,8.51482580646848)(27,9.0743456809288)(28,9.65581750452364)(29,10.0908640746084)(30,10.5542312301425)(31,10.8032779414289)(32,11.0218134989658)(33,11.1481613574289)(34,11.1590634780121)(35,11.0969309006086)(36,11.0488464328203)(37,10.9398839376529)(38,10.8153638010869)(39,10.6789672680027)(40,10.522605771579)(41,10.437007986067)(42,10.346971334275)(43,10.3164769270621)(44,10.2596132884612)(45,10.2613775604194)(46,10.307606266674)(47,10.3470802580436)(48,10.4128243252611)(49,10.5097872544925)(50,10.6150688548185)(51,10.747420512417)(52,10.8721706359712)(53,10.9918395744183)(54,11.1659828437591)(55,11.3464117415315)(56,11.4927992137989)(57,11.6480059167353)(58,11.8344997928029)(59,11.9798059232051)(60,12.1813457957004) 
};
\addlegendentry{1SNR=0~dB};
\end{axis}
\end{tikzpicture}\caption{Access latency for CSMUD.}
	\label{fig:csmudAL}
\end{figure}
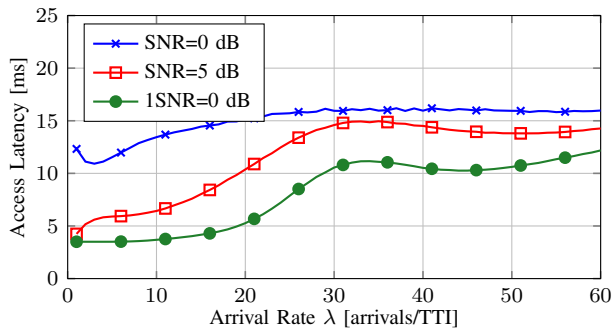

To evaluate \tblue{CSMUD} with respect to the KPIs and assumptions outlined in section~\ref{sec:fantastic_5g_approach} the physical layer approach \tblue{CSMUD} was combined with a simple one-stage protocol with random backoff according to the parameters of table~\ref{tab:params}. Each active user transmits its data in the current \tblue{TTI} and repeats this transmission in case of failure up to four times.
A combined MAC and PHY numerical simulation was conducted over $10^4$ trails including the full physical layer processing (encoding, modulation, channel estimation, multi-user detection, demodulation, decoding) with BPSK, a $[5;7]_8$ convolutional code, least squares multi-user equalization and BCJR decoding.
The activity and channel estimation step is achieved by the group orthogonal matching pursuit algorithm (GOMP). The spreading factor is $N_S=32$ and up to $K=64$ unique spreading sequences / preambles are considered.
The traffic model follows a Poisson arrival process with \tblue{an arrival} rate as shown on the x-axis of Fig.~\ref{fig:csmudPT} and \ref{fig:csmudAL}. Both KPIs depend on the signal-to-noise ratio (SNR), which is here assumed to \tblue{be either 0, 5 or 10~dB} and identical for all users. Hence, we implicitly assume some form of open-loop power control with idealized conditions.

Fig.~\ref{fig:csmudPT} depicts the protocol throughput of the \tblue{CSMUD} scheme which clearly shows a nearly linear scaling with increased arrival rate for the 10~dB case up to \tblue{an arrival rate of 16} at which the probability of successfully detecting and decoding a user begins to \tblue{decline} due to the interference level and a strong increase in retransmissions. \tblue{Surprisingly, beyond $\lambda=32$ the throughput increases again. This can be explained by the performance of the GOMP algorithm that is employed to solve \eqref{eq:csdet}. Up to half of the available sequences $K$ the detection performance declines because the number of non-zeros to be estimated increases up to the maximum potential for errors at exactly $K/2$. Due to the chosen stopping criteria the estimated channels $\hat{\mathbf{h}}$ can be 100\% wrong, i.e. all active users are estimated as inactive (missed detections) and all inactive users are estimated as active (false alarm). Beyond $K/2$, however, detection performance increases again with the decreasing number of zeros in the estimated vector. This is finally limited by the least squares performance of a two times overloaded CDMA system at 64 active users. Naturally, this behavior is also reflected for the lower SNRs of 5~dB and 0~dB with overall decreased performance.} Note, that the CSMUD approach used here does not exploit retransmission in any way. A combined decoding approach like presented in Section~\ref{sub:coded_random_access_with_PLNC} can strongly improve performance in cases where single slots are overloaded. However, this is highly dependent on the specific parameters of the system \cite{ji2017}. Furthermore, a comparison with the results presented in Section~\ref{sub:one_stage_vs_two_stages_access_protocols} indicates that the numerical simulations presented here behave differently than the pure MAC performance given orthogonal resources. Especially, the slope is lower, but the performance peak also occures later and seems broader hinting at a more robust behavior.

Fig.~\ref{fig:csmudAL} presents the access latency which is very low \tblue{for all} presented working points and shows much lower overall latencies than other schemes. Obviously, a single transmission is sufficient most of the time for 10~dB, which is increased with lower SNR and higher arrival rates. \tblue{The discussed GOMP behavior does not influence the latency as strongly but leads to small variations around the maximum latency. The access} latency is much lower than for example using signature based access or the frame focused PLNC enhanced scheme described in Section~\ref{sub:coded_random_access_with_PLNC}. This is easily explained by the fact that the both have to aggregate multiple TTIs to facilitate a successful access compared to the setup used here. 

\subsection{Coded Random Access with Physical Layer Network Coding \tblue{(CRAPLNC)}} 
\label{sub:coded_random_access_with_PLNC}
This proposal is inspired by a random access scheme aiming at reduced signalling. More specifically it considers physical layer techniques aiming to increase collision resolution through advanced receiver processing, and their integration with the MAC protocol. It targets one-stage protocols, although the PHY layer solution can be also exploited in two-stage protocols by allowing more than one packet transmission per radio resource block and resolving collisions through advanced receivers.
The solution falls under the category of coded random access \cite{PSLP2015}, where features of channel coding are exploited both at the slot and frame level. In particular, the scheme partially presented in \cite{CP14,PNC14} is extended for massive access, with emphasis on the transmission of short packets. The proposed scheme assumes a minimum coordination that ensures packet synchronization. The strategy focuses on collision resolution in a frame slotted ALOHA medium access scheme, where users are granted certain level of redundancy per transmission attempt. It exploits two features of coded schemes: the first one relates to the property that in the finite-field $\F{2}$, although the individual messages can not be correctly decoded, a linear combination of them (the bitwise XOR of a set of messages) may be. This property led to the so-called compute-and-forward \cite{NG2011,GGW2015}, which proved \tblue{achievable} gains, from an information-theory point of view. The second one exploits the increase in the diversity order of a linear system of equations if it is defined over an extended Galois Field $\mathbb{F}_{q}$ with field order $q=2^{n}$.
\begin{figure}[t]
	\centering
	\includegraphics[width=0.95\columnwidth]{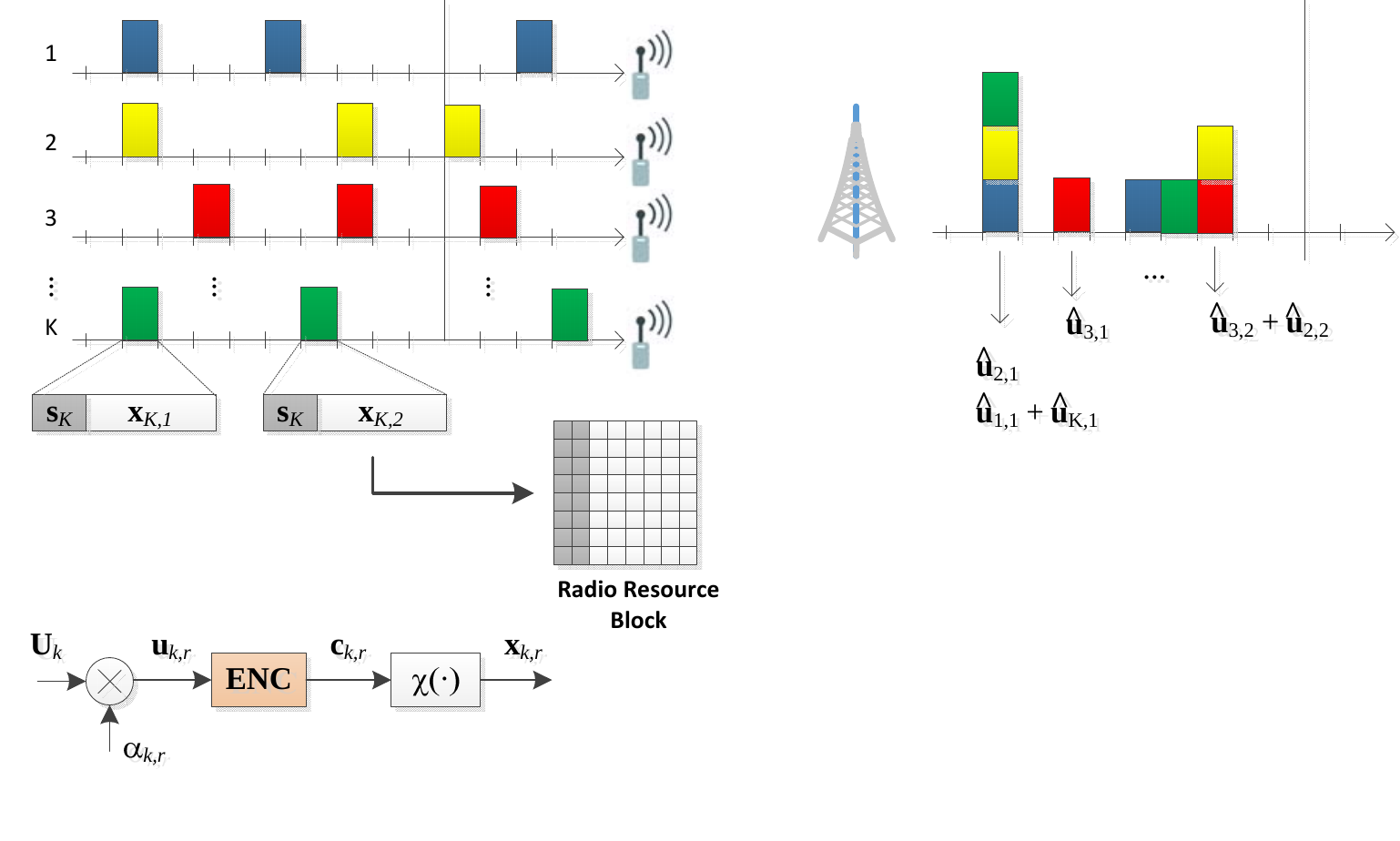}
	\caption{Coded Random Access, with PLNC and extended Galois field precoding.}
	\label{fig:craplncIlust}
\end{figure}
The multiple access scheme operates as follows:
\begin{itemize}
  \item Given a frame size of $S$ slots, users pick at random the slot positions where they will attempt transmission of each redundancy packet\footnote{In general, redundancy packets are not the same, they correspond to different codewords per user message.}. In principle the scheme can operate with a different level of redundancy $R$ and distribution. Although the illustrative example in Fig. \ref{fig:craplncIlust}, sets $R=2$ for all users, the scheme can be combined with optimized distributions.
  \item Each message, previous to channel encoding and modulation\footnote{The same channel code and modulation among users is assumed.}, allows for a linear pre-coding, which consist in a symbol-wise multiplication in the extended Galois Field $\F{q}$, i.e.~$\mathbf{u}_{\mathbb{F}_q}(m) = \alpha_r \times \mathbf{U}_{\mathbb{F}_q}(m)$ where $\mathbf{U}_{\mathbb{F}_q}(m)$ denotes the $m$-th symbol of the non-binary representation of the binary message $\mathbf{U}$. Pre-coding coefficients $\alpha_r\in
  \mathbb{F}_q$ are generated randomly\footnote{Note that the system can be configured to include no pre-coding, $\alpha_r\in \mathbb{F}_2$}.
  \item User detection and channel estimation is enabled by means of a preamble including the user signature and small overhead for identification of pre-coding coefficients.
  \item At the receiver side, for each slot, the receiver performs user detection and channel estimation, followed by the channel decoding stage. Each decoded message or linear combination (in $\F{2}$), generates a new row at the frame matrix $\mathbf{A}\in\F{q}$. If $\mathbf{A}$ is full rank, collisions can be resolved without the need for having one singleton packet.
\end{itemize}

For two-stage protocols, only the PHY layer component is used, applying advanced decoding to the reception of the data transmission stage. That is, the data transmission stage can be modified to allow several users to transmit their messages over the same physical resources. It only transmits the payload data since in this scenario, the receiver knows which users are transmitting and simply takes advantage of the increased capture probability provided by the advanced decoding scheme.

Relevant aspects of the scheme rely on the detection, channel estimation and decoding algorithms applied to the received signal within a single slot. In particular, for the detection of colliding users and channel estimation in one-stage protocols, we resort to a CSMUD algorithm, as introduced in section~\ref{sub:compressive_sensing_multi_user_detection} (see also  \cite{BSD2013}). More specifically, we consider channels with no delay spread and, thus, the simple CSMUD form in~\eqref{eq:csdet} is sufficient. Note that in the case where a packet fits a single radio resource block, as it is the case for the minimum allocation size of 1 PRB = 180~kHz $\times$ 1~ms (see Table I), the channel can be assumed constant. At the receiver side, advanced decoding (joint decoder and the "seek-and-decode" principle) is implemented independently at each slot, applied after standard SIC fails to decode any more messages, thus reducing complexity. Final decision decoding is made at the end of the frame ($S$ slots), although variants to the scheme could allow faster acknowledgments as soon as individual messages are correctly decoded at each slot. Results are shown in Figs.~\ref{fig:craplnc_throughput} and~\ref{fig:craplnc_latency} for very short codes (i.e.~binary LDPC with codeword length of 164 coded symbols) and system parameters defined in Table~\ref{tab:params} under block fading channels, shows relevant throughput gains against benchmark (slotted ALOHA) for moderate/high loads, even with no pre-coding. Results are also encouraging in terms of robustness against channel estimation errors, and user misdetection. We shall remark that simulation results include full physical layer implementation (multi-user detection, channel estimation and decoding) over the medium access control (for a frame size of $S=10$ slots). Further details can be found in~\cite{FANTASTIC-5GD4.2} including additional KPIs.
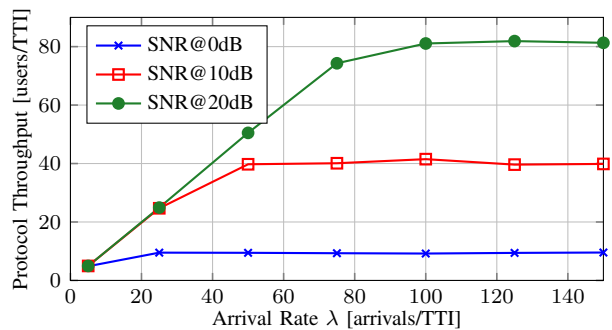
\begin{figure}
	\centering
%
%
%
%
\begin{tikzpicture}
\begin{axis}[%
small,
width=0.8\columnwidth,
height= 3.5cm,
scale only axis,
  try min ticks=6,
  max space between ticks=30pt,
  x tick label style={
    /pgf/number format/.cd,
    fixed,
    precision=2
  },
xmajorgrids,ymajorgrids,
xmin=0, xmax=3*50,
xlabel={Arrival Rate $\lambda$ [arrivals/TTI]},
ymin=0, ymax = 90,
ylabel={Protocol Throughput [users/TTI]},
legend cell align={left},
label style ={font=\footnotesize},
legend pos=north west,
legend style={font=\footnotesize}]


\addplot [
mark=x,
color=blue,
solid,
x filter/.code={\pgfmathparse{\pgfmathresult*50}\pgfmathresult}]
coordinates{
    (0.1,4.860)
    (0.5,9.510)
    (1.0,9.450)
    (1.5,9.315)
    (2.0,9.205)
    (2.5,9.425)
    (3.0,9.550)
};
\addlegendentry{SNR@0dB};
\addplot [mark=square,
color=red,
solid,
x filter/.code={\pgfmathparse{\pgfmathresult*50}\pgfmathresult}]
coordinates{
    (0.1,4.9611)
    (0.5,24.6887)
    (1,39.7471)
    (1.5,40.0973)
    (2.0,41.4981)
    (2.5,39.6304)
    (3,39.8638)
};
\addlegendentry{SNR@10dB};
\addplot [mark=*,
color=darkgreen,
solid,
x filter/.code={\pgfmathparse{\pgfmathresult*50}\pgfmathresult}]
coordinates{
    (0.1,4.9611)
    (0.5,24.9222)
    (1,50.4864)
    (1.5,74.2996)
    (2.0,81.0700)
    (2.5,81.8872)
    (3,81.3035)
};
\addlegendentry{SNR@20dB};
\end{axis}
\end{tikzpicture}
	\caption{Throughput performance of CRAPLNC massive access scheme for several SNRs.}
	\label{fig:craplnc_throughput}
\end{figure}
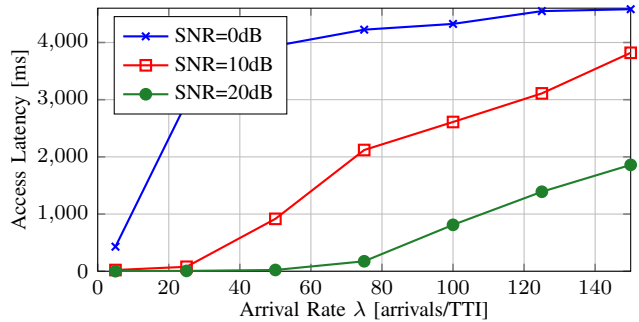
\begin{figure}
	\centering
%
%
%
%

\begin{tikzpicture}
\begin{axis}[%
small,
width=0.8\columnwidth,
height= 3.5cm,
scale only axis,
  try min ticks=6,
  max space between ticks=30pt,
  x tick label style={
    /pgf/number format/.cd,
    fixed,
    precision=2
  },
  y tick label style={
    /pgf/number format/.cd,
    fixed,
    precision=0
  },
xmajorgrids,ymajorgrids,
xmin=0, xmax=3*50,
xlabel={Arrival Rate $\lambda$ [arrivals/TTI]},
ymin=0, ymax = 4600,
ylabel={Access Latency [ms]},
label style ={font=\footnotesize},
legend pos= north west,
legend style={font=\footnotesize}]

\addplot [
mark=x,
color=blue,
solid,
x filter/.code={\pgfmathparse{\pgfmathresult*50}\pgfmathresult}]
coordinates{
(0.100000,429.928058)
(0.500000,2952.718191)
(1.000000,3940.603175)
(1.500000,4224.417606)
(2.000000,4325.086862)
(2.500000,4548.822281)
(3.000000,4582.073298)
};
\addlegendentry{SNR=0dB};
\addplot [mark=square,
color=red,
solid,
x filter/.code={\pgfmathparse{\pgfmathresult*50}\pgfmathresult}]
coordinates{
  (0.1,21.4)
    (0.5,77.9)
    (1,916.3)
    (1.5,2120)
    (2.0,2610)
    (2.5,3110)
    (3,3820)
};
\addlegendentry{SNR=10dB};
\addplot [
mark=*,
color=darkgreen,
solid,
x filter/.code={\pgfmathparse{\pgfmathresult*50}\pgfmathresult}]
coordinates{
    (0.1,1.4)
    (0.5,6.1)
    (1,22.3)
    (1.5,174.2)
    (2.0,811.3)
    (2.5,1390)
    (3,1860)
};
\addlegendentry{SNR=20dB};
\end{axis}
\end{tikzpicture}
	\caption{Latency performance of CRAPLNC massive access scheme for several SNRs.}
	\label{fig:craplnc_latency}
\end{figure}
%


\subsection{Compressive Sensing Coded Random Access \tblue{(CCRA)}} 
\label{sub:compressive_sensing_coded_random_access}
\begin{figure}[th]
\begin{center}
\includegraphics[width=\columnwidth]{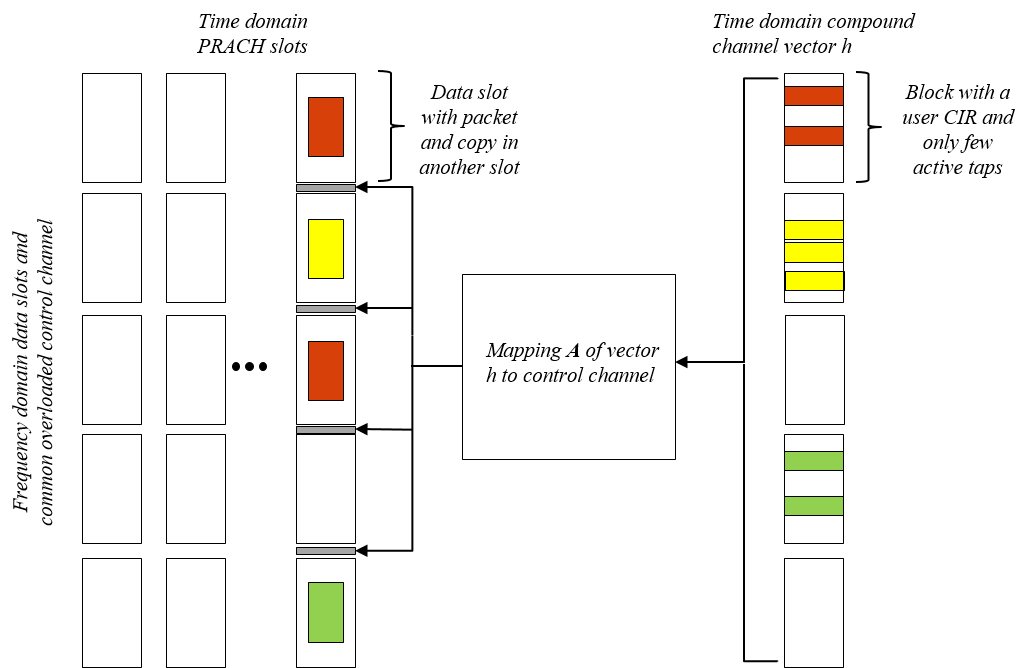}
\end{center}
\caption{Schematic of the CCRA scheme: sets are ... the common control channel}
\label{fig:ccra_scheme}
\end{figure}

Recent concepts combine advanced MAC protocols with \tblue{Compressive} Sensing (CS)
based multiuser detection \cite{Yi2014_GC,Wunder2015_ACCESS}. In this section,
we introduce a concept for sparse joint activity, channel and data detection
in the context of the Coded ALOHA (FDMA) protocol which we call
\emph{Compressive Coded Random Access} (CCRA) extending the work in
\cite{Yi2014_GC,Wunder2014_ICC,Wunder2015_GC,Wunder2015_ASILOMAR}. We will
argue that a simple sparse activity and data detection is not sufficient (as
many papers do) because control resources are in the order of the data. In
addition, we will 1) improve on the performance of such protocols in terms of
the reduction of resources required for the user activity, channel estimation
and data detection 2) achieve the required channel estimation quality for the
successive interference cancellation procedured required in coded ALOHA and CCRA.

Let us assume for simplicity a single time slot and an OFDM system with $n$
subcarriers. This is easily generalized to the case where there are multiple
time slots, notably, within the coherence time so that channels are constant
over these slots. Let $p_{i}\in\mathbb{C}^{n}$ be some signature from a given
set $\mathcal{P}\subset\mathbb{C}^{n}$ and $x_{i}\in\mathcal{X}^{n}$ be an
unknown (uncoded) data sequence (e.g. BPSK) from the modulation alphabet
$\mathcal{X}^{n}$ both for the $i$-th user with $i\in\{1,...,u\}$ and $u$ is
the (fixed) maximum set of users in the systems. Note that in our system $n$
is a very large number, e.g. 24k. Due to the random zero-mean nature of
$x_{i}$ we have $\frac{1}{n}E\lVert p_{i}+x_{i}\rVert_{\ell_{2}}^{2}=1$, i.e.~the total (normalized) transmit power is unity. Provided user $i$ is active,
we set:
\begin{equation}
\alpha:=\frac{1}{n}\lVert p_{i}\rVert_{\ell_{2}}^{2}\quad\text{and}\quad
\alpha^{\prime}:=1-\alpha=\frac{1}{n}E\lVert x_{i}\rVert_{\ell_{2}}^{2}%
\end{equation}
Hence, the control signalling fraction of the power is $\alpha$. If a user is
not active then we set both $p_{i}=x_{i}=0$, i.e.~either a user is active and
seeks to transmit data or it is inactive. Whether or wether not a user is
active depends on the traffic model and is discussed below.

Let $h_{i}\in\mathbb{C}^{s}$ denotes the sampled channel impulse response
(CIR) of user $i$ where $s\ll n$ is the length of the cyclic prefix (further
structural assumptions on $h_{i}$ are also discussed below). Let $[h_{i}%
,0]\in\mathbb{C}^{n}$ denote the zero-padded CIR. The received signal
$y\in\mathbb{C}^{n}$ is then:
\begin{align}
y &  =\sum_{i=0}^{u-1}\text{circ}([h_{i},0])(p_{i}+x_{i})+e\\
y_{\mathcal{B}} &  =\Phi_{\mathcal{B}}y
\end{align}
Here, $\text{circ}([h_{i},0])\in\mathbb{C}^{n}$ is the circulant matrix with
$[h_{i},0]$ in its first column. The AWGN is denoted as $e\sim
\mathcal{CN}(0,\sigma^{2})\in\mathbb{C}^{n}$, i.e.~$E(ee^{\ast})=\sigma
^{2}I_{n}$. $\Phi_{\mathcal{B}}$ denotes some measurement matrix (to be
specified) where the active rows indices are collected in $\mathcal{B}$ with
cardinality $m$. Typically, $\mathcal{B}$ refers to some set of subcarriers in
case of Fourier (FFT) measurements ($\Phi$ is orthonormal matrix) but, mainly
for analytical purposes, also Gaussian measurements are \tblue{considered} ($\Phi$ is
\emph{not} orthonormal matrix).

The key idea of CCRA scheme is that all users' preambles $\hat{p}_{i}\;\forall
i$ 'live' entirely in $\mathcal{B}$ while all data resides in the complement
$\mathcal{B}^{C}$, i.e.~formally supp$(\hat{p}_{i})\subseteq\mathcal{B}%
\;\forall i$, (hence, for orthonormal matrix $\Phi$ like FFT there is no
interference in between). We will call this a \emph{common overloaded control
channel} \cite{Wunder2015_GC} which is used for \emph{the user activity and
channel detection}. Since data resides only in $\mathcal{B}^{C}$ the entire
bandwidth $\mathcal{B}^{C}$ can be divided into $B$ frequency patterns. Each
pattern is uniquely addressed by the preamble and indicates where the data and
corresponding copies are placed. the scheme works as follows: if a user wants
to transmit a small data portion, the pilot/data ratio $\alpha$ is fixed and a
preamble is randomly selected from the entire set. The signature determines
where (and how many of) the several copies in the $B$ available frequency
slots are placed which are processed in a specific way (see below). Such
copies can greatly increase the utilization and capacity of the traditional
e.g. ALOHA schemes and which is used for \emph{the data detection}. An
illustration of the scheme is in Fig.~\ref{fig:ccra_scheme}.

To derive a proper model for the user activity and channel detection, we can
stack the users as:%

\begin{equation}
y=D(p)h+C(h)x+e
\end{equation}
where $D(p):=[$circ$^{(s)}(p_{1}),\dots,$circ$^{(s)}(p_{u})]\in\mathbb{C}%
^{n\times us}$ and $C(h):=[$circ$^{(n)}([h_{1},0]),\dots,$circ$^{(n)}%
([h_{u},0])]\in\mathbb{C}^{n\times un}$ are the corresponding compound
matrices, respectively $p=[p_{1}^{T}\ p_{2}^{T}\ ...p_{u}^{T}]^{T}$ und
$h=[h_{1}^{T}\ h_{2}^{T}\ ...h_{u}^{T}]^{T}$ are the corresponding compound
vectors. In general, the measurement map is difficult to analyze since $D(p)$
depends on the specific design of the signatures $p_{i}$. One choice of
$\mathcal{P}$ that works for a small number of active users and $n\gg us$ is
as follows: We choose $p_{0}$ to be a sequence with unit power in frequency
domain, i.e.~such that (up to phases, which can be selected according to other
optimization criteria, e.g. PAPR):
\begin{equation}
|\left(  \hat{p}_{0}\right)  _{i}|=\left\{
\begin{array}
[c]{cc}%
\sqrt{\frac{n}{m}} & i\in\mathcal{B}\\
0 & \text{else}%
\end{array}
\right.
\end{equation}
where $\hat{p}_{0}:=Wp_{0}$ denotes the FFT transform of $p_{0}$. Since $n\geq
us$, the matrix $D(p)$ can be completely composed of cyclical shifts of the
sequence $p_{0}$, i.e.~$p_{1}=p_{0},\quad p_{2}=p_{1}^{\left(  s\right)
},\quad p_{3}=p_{2}^{\left(  s\right)  },\quad\ldots\ ,$where $p^{\left(
i\right)  }$ is the $i$ times cyclically shifted $p$. Hence, $D(p)$ is a
single circulant matrix, and, in this situation, we can show, that the control
channel is finally represented as:%
\[
y_{\mathcal{B}}=Ah+z,
\]
where $A$ is a subsampled $m\times us$ FFT matrix, which is normalized by a
factor of $\sqrt{1/m}$ and $z\sim\mathcal{CN}\left(  0,\frac{\sigma^{2}}%
{n}I_{m}\right)  $. Now, based on this \tblue{measurement model}, the most
important assumptions on the structure of $h$ are:

\begin{itemize}
\item Bounded support of $h_{i}$ (with high probability), i.e.~supp$(h_{i}%
)\leq s$ and $s\ll n$

\item Sparse user activity, i.e.~$k_{u}$ users out of $u$ are actually active

\item Sparsity of $h_{i}$, i.e.~$\lVert h_{i}\rVert_{l_{0}}\leq k_{s}$
\end{itemize}

Hence, classical sparsity of $h$ is $k:=k_{u}k_{s}$ and the typical arsenal of
CS algorithms can be used. In CCRA, though, we are exploiting block-column
sparsity: a $k$-sparse compound vector $h$ is so-called block-column sparse,
i.e.~$(k_{u},k_{s})$-sparse, if it consists of $k_{u}$ active blocks of length
$s$ each $k_{s}$-sparse. Block-column sparsity is exploited in the detection
of the activity and channel by a new algorithm called Hierarchical HTP
(HiHTP). HiHTP uses a so-called block-column tresholding operator
$L_{k_{s},k_{u}}(z)$. This operator can be efficiently calculated by selecting
the $k_{s}$ absolutely largest entries in each block and subsequently the
$k_{u}$ blocks that are largest in $\ell_{2}$-norm. The strategy of the HiHTP
algorithm is to use the thresholding operator $L_{k_{u},k_{s}}$ to iteratively
estimate the support of $h$ and subsequently solve the inverse problem
restricted to the support estimate. HiHTP comes with explicit recovery
guarantees while exploiting the specific structure of $h$, see
\cite{Wunder2016_ARXIV}.

The data detection algorithm can be seen as in instance of \emph{coded slotted
ALOHA} framework \cite{PSLP2015}, tuned to incorporate the particularities of
the physical layer addressed in the paper, as described in the previous
section. Specifically, the random access algorithms assumes that:

\begin{itemize}
\item the users are active in multiple combinations of time-frequency slots,
denoted simply as slots in further text,

\item the activity pattern, i.e.~the choice of the slots is random, according
to a predefined distribution,

\item every time a user is active, it sends a replica of packet, which
contains data,

\item each replica contains a pointer to all other replicas sent by the same user.
\end{itemize}

Obviously, due to the random nature of the choice of slots, the access point
(i.e.~the base station) observes idle slots (with no active user), singleton
slots (with a single active user) and collision slots (with multiple active
users). Using a compressive sensing receiver, the base station, decodes
individual users from non-idle slots, learns where the replicas have occurred,
removes (cancels) the replicas, and tries to decode new users from slots from
which replicas (i.e.~interfering users) have been cancelled. In this way, due
to the cancelling of replicas, the slots containing collisions that previously
may have not been decodable, can become decodable. This process is executed in
iterations, until there are no slots from which new users can be decoded. The
above described operation can be represented via graph. Analytical modeling of
the above is the main prerequisite to assess the performance of the random
access algorithm, which in turn, allows for the design of the probability
distribution that governs the choice slots, and which is typically optimized
to maximize the throughput, i.e.~the number of resolved packets per slot
\cite{PSLP2015}.

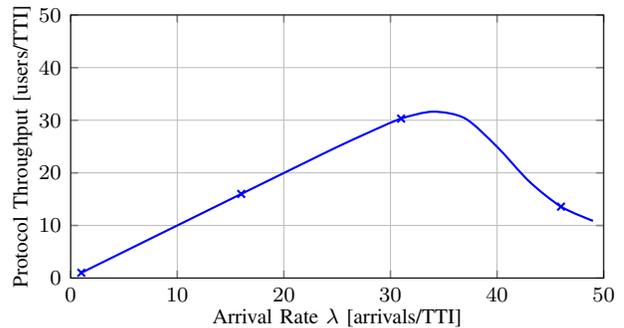
\begin{figure}[th]
	\begin{center}
%
%
%
%

\begin{tikzpicture}
\begin{axis}[%
small,
width=0.8\columnwidth,
height= 3.5cm,
scale only axis,
  try min ticks=6,
  max space between ticks=30pt,
  x tick label style={
    /pgf/number format/.cd,
    fixed,
    precision=2
  },
xmajorgrids,ymajorgrids,
xmin=0, xmax=50,
xlabel={Arrival Rate $\lambda$ [arrivals/TTI]},
ymin=0, ymax =50,
ylabel={Protocol Throughput [users/TTI]},
label style ={font=\footnotesize},
legend pos=north west,
legend style={font=\footnotesize}]

\addplot [smooth,mark=x,
mark repeat=5,
color=blue,
solid,
x filter/.code={\pgfmathparse{\pgfmathresult*50}\pgfmathresult},
y filter/.code={\pgfmathparse{\pgfmathresult*50}\pgfmathresult}]
coordinates{
(	0.02	,	0.02	)
(	0.08	,	0.08)
(	0.14	,	0.14)
(   0.2  ,    0.19992)
(	0.26	,	0.2598)
(	0.32	,	0.31972)
(	0.38	,	0.37946)
(  0.44  ,   0.43906)
(  0.5   ,    0.49836)
(0.56 , 0.55422)
(0.62 , 0.6061)
(0.68 , 0.63248)
(0.74 , 0.6069)
(0.8 , 0.49912)
(0.86 , 0.3679)
(0.92 , 0.27144)
(0.98 , 0.2175)
};

\end{axis}
\end{tikzpicture}
	\end{center}
	\caption{\tblue{Throughput performance of CCRA over arrival rate.}}
	\label{fig:T_G}
\end{figure}

\tblue{We follow the common simulation assumptions described in Table I. Note
that the pilot-to-data ratio is only 13\% so the overhead compared to
LTE-4G has significantly reduced (reported to be up to 2000\% in
\cite{Wunder2014_COMMAG}). For the CCRA throughput evaluation, we use
BPSK modulated subcarriers and successive interference
cancellation. Fig. \ref{fig:T_G} shows the throughput of actually
successfully recovered packets over different arrival rates using
\emph{at most} three replicas per packet (optimum results from testing
one to five copies). It can be seen that with three replicas the
performance is significant improved over, say, traditional slotted
ALOHA (SA) which achieves only max. 40\% normalized throughput (i.e.~20 user/TTI). While
not shown here in detail, we mention that BER performance for
detecting replicas at 15dB SNR is well below $10^{-1}$ even for those
with three-step interference cancellation detection procedure pointing
out the good channel estimation performance. Altogether, we conclude
that even for this challenging scenario the CCRA achieves a
significant throughput gain with reasonable BER performance per
detected and decoded packet and, at the same time, drastically reduces
the signalling overhead.}

\subsection{Slotted Compute and Forward (SCF)} 
\label{sub:slotted_compute_and_forward}
\begin{figure}[t]
\centering
\def\svgwidth{600pt}%
\scalebox{0.52}{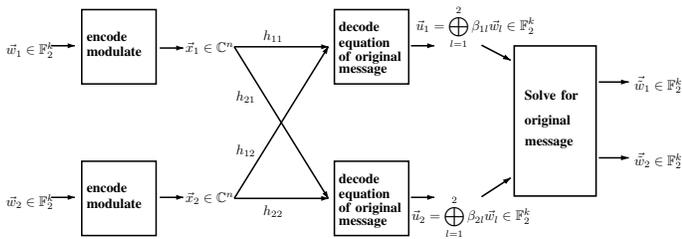}
\caption{\tblue{Toy example describing the main processing blocks for the 2 transmitters $\times$ 2 mini base stations case.}}
\label{fig:scheme}
\end{figure}

\tblue{The presented Slotted Compute-and-Forward (SCF) approach is a random access extension of the Compute-and-Forward (CF) relaying scheme introduced in \cite{Nazer:Gastpar:11}. The approach combines the concept of network densification with physical-layer network coding and a multicarrier transmission scheme (OFDM). Using linear codes it enables the network to exploit channel collisions~\cite{Goldenbaum:Stanczak:13a} by decoding linear combinations of the messages transmitted by different devices that access the channel simultaneously in the same frequency band. The scheme assumes a dense network infrastructure with a large number of MTC devices accessing the wireless channel, where each transmitter can be heard by multiple mini base stations. The data transmission is a two-hop communication with multiple mini base stations acting as relays. They receive individual superpositions of the sent signals, process, decode and forward them to the macro base station. The macro base station then estimates the transmitted messages over a finite field based on the received linear combinations. A simplified example is shown in Fig.~\ref{fig:scheme}.}

Let us  assume that the large set of uniformly distributed MTC devices $\mathcal{M}_{tot}$, with $M_{tot}:=
|\mathcal{M}_{tot}|$, is supported by a set of mini base stations $\mathcal{B}_{tot}$, which are connected to the macro base station through a wired or wireless communication. Each mini base station has only knowledge of its own channel coefficients, whereas the MTC devices \tblue{have no channel} state information.
Let $\mathcal{M}\subset\mathcal{M}_{tot}$, with $M:=
|\mathcal{M}|$, be a set of active MTC devices that can be heard by each mini base station $b\in\mathcal{B}$ of a predefined subset $ \mathcal{B}\subset \mathcal{B}_{tot}$ and $B:=
|\mathcal{B}|$. Note that, for simplicity, we have assumed here $B=M$. To increase robustness it is often reasonable to choose $B>M$ and solve instead the over-determined system \tblue{of} equations. Each device $m\in\mathcal{M} $, has a length-$k$ \tblue{complex} message $\ve{w}_m=(\ve{w}^R_m, \ve{w}^I_m)$, with $\ve{w}^R_m$ and $\ve{w}^I_m$ real, respectively \tblue{imaginary part} drawn from some finite field $\mathds{F}_p^k$\tblue{,} and maps its message to a length-$n$ codeword $\ve{x}_m\in\mathds{C}^n$ subject to an average power constraint $\frac{1}{n}\|\ve{x}_m\|^2\leq P$. We model the complex baseband signal $\ve{y}_b$ received by mini base station $b\in\mathcal{B}$ as %
\begin{equation} \ve{y}_b=\sum_{m\in\mathcal{M}}h_{bm}\ve{x}_m+\ve{z}_b\;,
\label{eq:receive_signal}%
\end{equation}%
where $\ve{z}_b\sim\mathcal{CN}(\ve{0},\ve{I}_n)$ denotes independent Gaussian noise of unit variance (per dimension) and $h_{bm}$ is the complex-valued channel coefficient between MTC device $m$ and base station~$b$.

\tblue{The mini base station $b$ performs rescaling and \textbf{integer--forcing} to obtain a noisy linear combination of the transmitted codewords with integer coefficients:}

\begin{equation*}
             \tilde{\ve y}_b=\alpha_b\ve y_b= \sum_{\mathclap{ m\in \mathcal{M}}} a_{bm}\ve{x}_m+ \underbrace{\sum_{\mathclap{ m\in \mathcal{M}}}(\alpha_b h_{bm}-a_{bm})\ve{x}_m +\alpha_b\ve z_b}_{\text{effective noise}},
\end{equation*}
\tblue{ and decode $\mathds{F}_p$ linear combinations  \[\ve u_b:=\bigoplus_{\mathclap{m\in \mathcal{M}}} \beta_{bm}\ve{w}_m \]  of messages $\ve{w}_m$ over $\mathds{F}_p$. The scaling factor $\alpha_b$ and the integer coefficients $a_{bm}$ are chosen such that the effective noise is minimized. The equation coefficients $\beta_{bm}$ satisfy $\beta_{bm}=[a_{bm}]\bmod{p} \in\mathds{F}_p$.}
Once the mini base stations have successfully decoded the linear equations they forward \tblue{these} along with the respective coefficients $\ve \beta_b = ( \beta_{b1} , \ldots , \beta_{bM} )$ to the macro base station. If the equation coefficients have been chosen such that the matrix $\mathds{B}:= (\ve{\beta}_1,\dots,\ve{\beta}_B)^T\in\mathds{F}_p^{B\times M}$ is invertible over $\mathds{F}_p$, the macro base station estimates the original messages by calculating \cite{Raceala-Motoc2016}. %
\begin{equation}\label{eq:B} 
(\hat{\ve{w}}_1,\dots,\hat{\ve{w}}_M)^T=\mathds{B}^{-1}(\hat{\ve{u}}_1,\dots,\hat{\ve{u}}_B)^T\;. \end{equation}%
\tblue{This approach makes the SCF solution especially suited for two-hop communication scenarios where the capacity limited second hop is a bottleneck in the transmission.}
For a transmission to be successful, the superposition of messages has to be successfully decoded, and $\mathds{B}$ must be invertible over $\mathds{F}_p $, meaning that the system of linear equations at the macro base station has to \tblue{be of} full rank. 
To reduce the probability of rank deficiency, each MTC device transmits the same message over several frequency slots. For our simulations we consider four slots.
We further allow for cooperation between mini base stations and macro base station for up to four colliding devices.

To analyze the end-to-end performance of the approach we consider a system which consists of the following blocks:  coding and modulation, resource allocation, transmission over the wireless channel, signal reception and processing at the mini base stations, forwarding to the macro base station, data aggregation at the macro base station.

We assume that channel estimation has been performed and all active devices have been identified \cite{Goldenbaum2016}.
For the slotted transmission synchronization within the guard interval is assumed. All nodes are equipped with a single antenna while all devices transmit at an equal rate. The messages are encoded using \tblue{an} LDPC channel code with a code rate of R=1/4. Each transmitter transmits complex messages of 128 bit, sending a total number of 256 information bit. The encoded data is modulated using a QPSK modulation alphabet. The channel is modeled as a four-tap block fading Rayleigh multipath channel. \tblue{For the sake of computational complexity we assume that no more than 9 devices collide on the same resource block at a given time. Note that the number of active devices during one time-slot can be much higher. Since each device transmits two independent messages over complex channels, up to 18 messages can collide.} 
 The traffic model follows a Poisson arrival process with an arrival rate $\lambda$ per time-slot.

Both KPIs, protocol throughput and access latency are highly dependent on the signal-to-noise ratio (SNR).

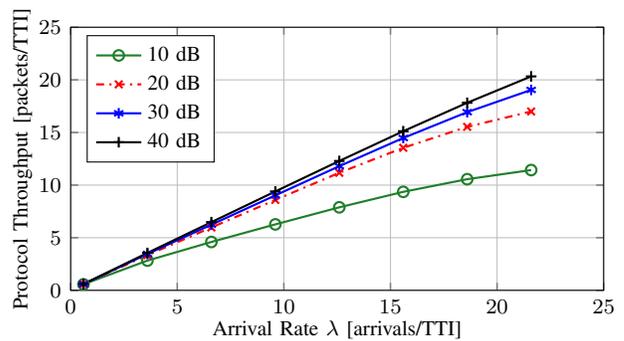
\begin{figure}
	\centering
%
\definecolor{mycolor1}{rgb}{1.00000,0.00000,1.00000}%
%
%

\begin{tikzpicture}
\begin{axis}[%
small,
width=0.8\columnwidth,
height= 3.5cm,
scale only axis,
  try min ticks=6,
  max space between ticks=30pt,
  x tick label style={
    /pgf/number format/.cd,
    fixed,
    precision=2
  },
xmajorgrids,ymajorgrids,
xmin=0, xmax=25,
xlabel={Arrival Rate $\lambda$ [arrivals/TTI]},
ymin=0, ymax =25,
ylabel={Protocol Throughput [packets/TTI]},
label style ={font=\footnotesize},
legend pos=north west,
legend style={font=\footnotesize}]

\addplot [color=darkgreen,solid,mark=o,mark options={solid}]
  table[row sep=crcr]{%
0.6	0.571863064331308\\
3.6	2.82270101513083\\
6.6	4.59499258709603\\
9.6	6.26539684589511\\
12.6	7.88686768570166\\
15.6	9.35869151524817\\
18.6	10.5605007489839\\
21.6	11.4278380810485\\
};
\addlegendentry{10 dB};

\addplot [color=red,dashdotted,mark=x,mark options={solid}]
  table[row sep=crcr]{%
0.6	0.590654268203972\\
3.6	3.34649541933046\\
6.6	5.97158538110741\\
9.6	8.59388800200693\\
12.6	11.1694173951879\\
15.6	13.543432012763\\
18.6	15.533181538491\\
21.6	16.9971811106993\\
};
\addlegendentry{20 dB};

\addplot [color=blue,solid,mark=asterisk,mark options={solid}]
  table[row sep=crcr]{%
0.6	0.594605492268646\\
3.6	3.45369953504007\\
6.6	6.23761241534761\\
9.6	9.0243870817167\\
12.6	11.7986498759187\\
15.6	14.4753547766915\\
18.6	16.9322349970183\\
21.6	19.045671574492\\
};
\addlegendentry{30 dB};

\addplot [color=black,solid,mark=+,mark options={solid}]
  table[row sep=crcr]{%
0.6	0.598217102090463\\
3.6	3.55160173050569\\
6.6	6.47810870330073\\
9.6	9.3962809735754\\
12.6	12.2905492356236\\
15.6	15.1197481121675\\
18.6	17.823219240262\\
21.6	20.3287820683662\\
};
\addlegendentry{40 dB};

\end{axis}
\end{tikzpicture}%
	\caption{\tblue{Throughput performance of SCF massive access scheme for several SNRs of the first hop.}}
	\label{fig:THP}
\end{figure}

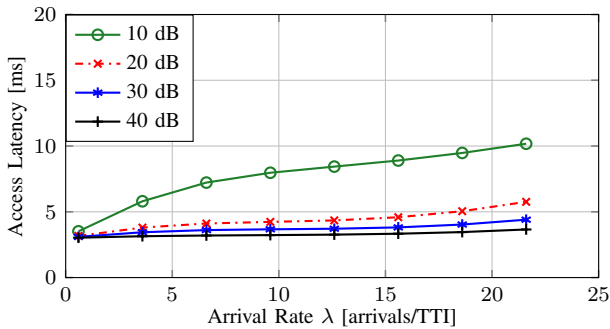
\begin{figure}
	\centering
%
\definecolor{mycolor1}{rgb}{1.00000,0.00000,1.00000}%
%
%
\begin{tikzpicture}
\begin{axis}[%
small,
width=0.8\columnwidth,
height= 3.5cm,
scale only axis,
  try min ticks=6,
  max space between ticks=30pt,
  x tick label style={
    /pgf/number format/.cd,
    fixed,
    precision=2
  },
xmajorgrids,ymajorgrids,
xmin=0, xmax=25,
xlabel={Arrival Rate $\lambda$ [arrivals/TTI]},
ymin=0, ymax =20,
ylabel={Access Latency [ms]},
label style ={font=\footnotesize},
legend cell align=right,
legend style={at={(0,1)}, anchor=north west,font=\footnotesize}]

\addplot [color=darkgreen,solid,mark=o,mark options={solid}]
  table[row sep=crcr]{%
0.6	3.51642025141532\\
3.6	5.79994671386998\\
6.6	7.22084666402812\\
9.6	7.96791212750626\\
12.6	8.43604796655758\\
15.6	8.89804830737696\\
18.6	9.47449431535092\\
21.6	10.1736143490695\\
};
\addlegendentry{10 dB};

\addplot [color=red,dashdotted,mark=x,mark options={solid}]
  table[row sep=crcr]{%
0.6	3.16638769885225\\
3.6	3.79983533977519\\
6.6	4.11095768881726\\
9.6	4.23371750822438\\
12.6	4.34517319909173\\
15.6	4.58636282278205\\
18.6	5.04446412359101\\
21.6	5.75765183089097\\
};
\addlegendentry{20 dB};

\addplot [color=blue,solid,mark=asterisk,mark options={solid}]
  table[row sep=crcr]{%
0.6	3.09526008207225\\
3.6	3.44467077124641\\
6.6	3.60963849346576\\
9.6	3.66919097872553\\
12.6	3.71246044693295\\
15.6	3.81465045255974\\
18.6	4.03150576376068\\
21.6	4.40005009777351\\
};
\addlegendentry{30 dB};

\addplot [color=black,solid,mark=+,mark options={solid}]
  table[row sep=crcr]{%
0.6	3.03129369927044\\
3.6	3.14308388073374\\
6.6	3.19756189464783\\
9.6	3.22764117069652\\
12.6	3.26435951029771\\
15.6	3.33349616808088\\
18.6	3.45753225892487\\
21.6	3.65616408992244\\
};
\addlegendentry{40 dB};

\end{axis}
\end{tikzpicture}%
	\caption{\tblue{Access Latency of SCF massive access scheme for several SNRs of the first hop.} }
	\label{fig:AL}
\end{figure}

\tblue{The throughput, shown in Figure \ref{fig:THP}, is defined as the mean number of successfully transmitted messages for a certain arrival rate $\lambda$. No retransmissions have been considered in the simulations when determining the protocol throughput. 
Since the macro base station is still able to decode, packets are not discarded when two or more collisions occur, leading to a high throughput even for the lower SNR region. In the considered setup the throughput does not improve significantly for SNR values above 20dB.
In order to determine the access latency, depicted in Figure \ref{fig:AL},  we combined the SCF physical layer approach with a random backoff protocol. Transmission is repeated in case of failure until a successful transmission or until the maximum number of retransmissions is reached. 
If a random access is successful at the first attempt, the expected latency includes the wake-up time and the time to perform a successful random access. If the random access is successful at a later attempt, the access latency includes the latency caused by the unsuccessful attempts prior to the successful transmission, the back-off time between retransmissions and the latency of the last successful random access.}
		
\tblue{Since users manage to transmit their messages on average in one or two transmissions, the SCF access latency can be kept very low.}



\subsection{Massive MIMO} 
\label{sub:massive_mimo}
We now consider a massive access solution that takes advantage of the Massive MIMO capabilities, where the base station of a massive MIMO system is equipped with a very large number of antennas and 
can create a very large number of spatial Degrees of Freedom (DoF) under favourable propagation conditions. 
Those DoFs are naturally suited to efficiently serve  a very large number of devices such as in machine-type communications, not only by spatially multiplexing a dense crowd of devices but also by improving contention resolution in resource access. 
\tblue{We target a multiple antenna system at legacy frequency band (below 6GHz) where the devices are assumed to have a small number of antennas due to their size. The use of a larger number of antennas at the devices is in principle possible at millimeter-wave bands. However, the cost of devices equipped with multiple antennas and beamforming capabilities at those bands  is currently a limitation in MTC applications. }

This solution addresses two important aspects in machine-type communications: acquisition of Channel State Information (CSI) and data communications for uplink traffic. 
CSI is estimated at the BS based on training via pilot sequences. The pilot sequences available are assumed to be mutually {\it orthogonal}.
For UL machine-type traffic, CSI estimation suffers from two fundamental limits. 
{First}, the duration of pilot sequences is limited by the (time-frequency) coherence interval of the channel, as well as the transmit power of the device. 
For orthogonal pilot sequences and in crowd scenarios, it means that the number of sequences could be in severe shortage. 
Therefore,  allocation policy of the pilot sequences becomes a central question. 
{Second},  the data traffic is intermittent and only a subset of the devices is active simultaneously. Hence a fixed pilot allocation to all the devices in the system would be highly inefficient. Pilot allocation has rather to adapt and scale with the traffic activity pattern and not to the actual number of devices present in the system.
A natural choice is to decentralize pilot access to the devices and make it random. 

Random access to pilot sequence leads to pilot collision, also known as \emph{pilot contamination}. Pilot contamination is a major impairment in massive MIMO system: when used for data decoding, contaminated channel estimates lead to interference that can be significant. 
The basic idea of the proposed joint pilot and data access is to randomize the effect of pilot contamination over multiple transmission slots, so that the 
the effect of contamination-induced interference is averaged out and becomes predictable. 
Related work can be found \cite{deCarvalho2016a,Sorensen2016a, Bjornson2017a}.

\begin{figure}[!t]
\centering
\includegraphics[width=8cm]{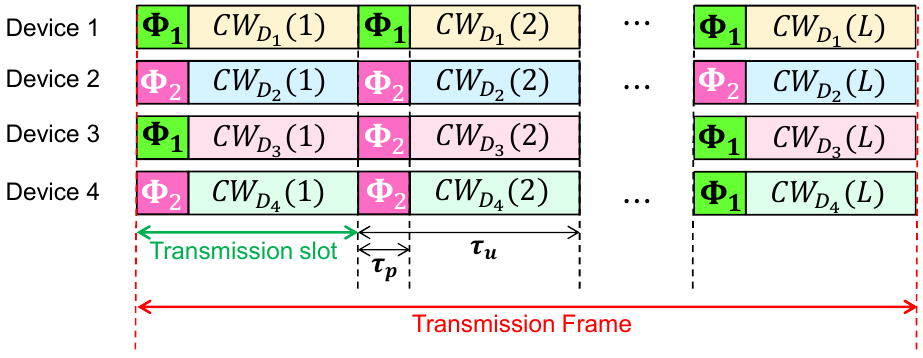} 
\caption{Illustration of the transmission frame with four active devices $\{D_1,D_2, D_3,D_4\}$ and two mutually orthogonal pilot sequences  
$\{\boldsymbol{\Phi}_1, \boldsymbol{\Phi}_2\}$.}
\label{fig:Model3}
\end{figure}

Uplink transmission is organized into transmission frames made out of multiple transmission slots. 
A transmission slot is a time-frequency unit where the channel can be approximated as constant. 
Fig.~\ref{fig:Model3} depicts a simplified example with four active devices and two orthogonal pilot sequences, where $\tau_u$ is the duration of
a transmission slot and $\tau_p$ is the duration of the pilot sequences in symbols.  
A block fading model is assumed, with independent realization in each slot and for each device. 

A device with data to transmit waits for the start of a new  transmission frame. 
Each active device encodes its data into one codeword that is divided into multiple parts and
transmits one pilot sequence followed by one part of  the codeword within a transmission slot. The pilot sequence serves to estimate the channel that is then used for soft decoding of the associated codeword portion.
Within a transmission frame, a number of $K_a$ devices are active out of a total number of $K$ devices. The activation probability of a device is $p_a$.

In order to randomize the effect of pilot contamination, \emph{pilot hopping} is performed. In each transmission slot, each active device 
 selects one pilot sequence from the set of orthogonal pilot sequences according to a pseudo-random pilot-hopping pattern that is unique to the device. Hence, in each transmission slot and for one given device, 
contamination-induced interference comes from different sets of devices. 
The  codeword of the device experiences all possible contamination events from the $K_a$ active devices, provided that the number of transmission slots duration is sufficiently long. Likewise, for an asymptotic large number of transmission slots, the additive noise at the BS and fading is averaged out. 
{
Under those asymptotic conditions, a maximal achievable rate per device can be defined within each transmission frame. 
Achieving this rate assumes the following features: a) estimation of the number of active users at the BS, 
b) estimation of the average channel energy per device at the BS and at the device, c) BS broadcasts the rate associated to each value of   average channel energy. 
With conditions a) and b), the BS computes a maximal achievable rate per device. 
The BS broadcasts both  the channel energy and its associate rate for each active device. 
As the device itself knows its channel energy, it can associate its assigned rate. }

For each transmission slot, the following steps are performed. 
First, the BS detects which pilot sequences  are in use. 
This is done by correlating the received signal  with each sequence available. 
{The pilot detection outcomes are buffered in order to be utilized for device activity detection.}
Second, for each pilot sequence detected, the corresponding channel estimate  is computed. In this work, MMSE channel estimation is performed. When there is pilot collision, channel estimation is contaminated. 
Third, for each pilot sequence detected, a multiple antenna processing based on the channel estimate is applied to the data symbols  in the slot and its  output is buffered along with its associated pilot index. In this work, Maximum Ratio Combining  (MRC) is utilized.  

A unique pseudorandom pilot-hopping pattern is assigned to each device. The pilot-hopping patterns are known at the BS  and serve for device identification at the BS. In order to detect the transmitting devices, the BS combines the pilot sequence detection outcomes from the slots that follow the pattern. 
Based on the identifying pilot-hopping patterns, the BS identifies which MRC outputs to combine to decode the data of each transmitting device.  

Our main performance metric is the system uplink sum rate. It is the sum rate per transmission frame averaged over the activation statistics of the device population. 
We work on  an approximation of the uplink sum rate ${\cal R}$ that is tight thanks to channel hardening and when the total number
of devices is large. This metric 
depends \tblue{on} the total number of BS antennas $M$
and the number of pilot sequences, $\tau_p$: the larger those quantities, the more devices can be multiplexed. 
Bound ${\cal R}$ is also a function of the device activation probability, $p_a$.  
To maximize the sum rate, one can optimize $p_a$ and $\tau_p$. 
When the number of antennas $M$ and the duration of transmission slot $\tau_u$ are of the same order, the sum rate  scales of 
$\sqrt{M \tau_u}$. Heuristic solutions indicate that one third of the transmission slot should be devoted to training while the average number of active devices should be of the order of $\sqrt{M \tau_u}$.

\begin{figure}
	\centering
%
%
%
%

\begin{tikzpicture}
\begin{axis}[%
small,
width=0.8\columnwidth,
height= 3.5cm,
scale only axis,
  try min ticks=6,
  max space between ticks=30pt,
  x tick label style={
    /pgf/number format/.cd,
    fixed,
    precision=2
  },
xmajorgrids,ymajorgrids,
xlabel={Arrival Rate $\lambda$ [arrivals/TTI]},
xmin=0, xmax=15,
ylabel={Protocol Throughput [users/TTI]},
ymin=0, ymax =20,
label style ={font=\footnotesize},
legend pos=north west,
legend style={font=\footnotesize}]

\addplot [mark=x,
color=blue,
solid,
x filter/.code={\pgfmathparse{\pgfmathresult+0.3}}]
table[row sep=crcr]{%
0.42671307006327	0.42671307006327\\
5.40469153146616	5.40469153146616\\
};
\addlegendentry{\# antennas M = 100}

\addplot [
mark=square,
mark repeat=5,
color=red,
solid,
x filter/.code={\pgfmathparse{\pgfmathresult}}] 
table[row sep=crcr]{%
0.565287620106576	0.565287620106576\\
9.23970893653913	9.23970893653913\\
};
\addlegendentry{\# antennas M = 200}

\addplot [
mark=*,
mark repeat=5,
color=darkgreen,
solid,
x filter/.code={\pgfmathparse{\pgfmathresult-0.3}}]
table[row sep=crcr]{%
0.711082843800288	0.711082843800288\\
14.3103625235142	14.3103625235142\\
};
\addlegendentry{\# antennas M = 400}

\end{axis}
\end{tikzpicture}
	\caption{Protocol throughout as a function of the arrival rate for $K = 400$ and  for $M=100,200,400$.}
	\label{fig:Perf1}
\end{figure}
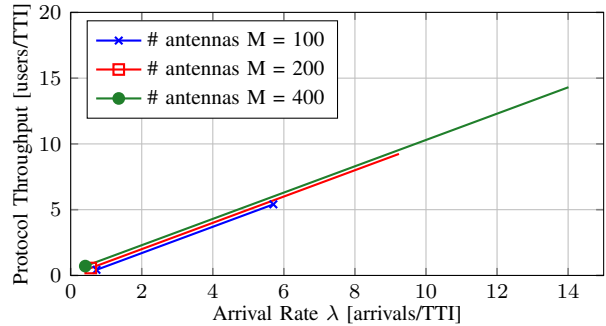

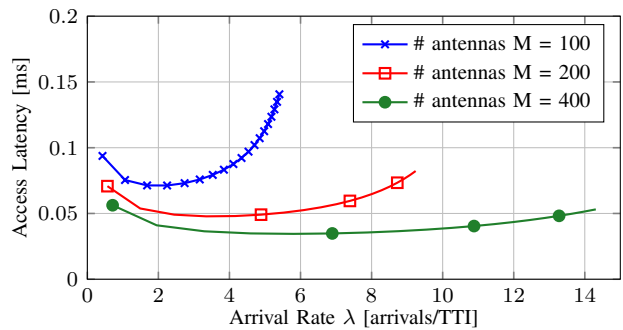
\begin{figure}
	\centering
%
%
%
%

\begin{tikzpicture}
\begin{axis}[%
small,
width=0.8\columnwidth,
height= 3.5cm,
scale only axis,
  try min ticks=6,
  max space between ticks=30pt,
  x tick label style={
    /pgf/number format/.cd,
    fixed,
    precision=2
  },
  y tick label style={
    /pgf/number format/.cd,
    fixed,
    precision=2
  },
xmajorgrids,ymajorgrids,
xmin=0, xmax=15,
xlabel={Arrival Rate $\lambda$ [arrivals/TTI]},
ymin=0, ymax = 0.2,
ylabel={Access Latency [ms]},
label style ={font=\footnotesize},
legend pos= north east,
legend style={font=\footnotesize}]

\addplot [mark=x,
color=blue,
solid]
table[row sep=crcr]{%
0.42671307006327	0.093739805049958\\
1.06048517491778	0.0754371696013594\\
1.68239899001486	0.071326718996033\\
2.24481794275082	0.0712752677858296\\
2.73791521091589	0.0730482811164543\\
3.16410657588877	0.0758507952383323\\
3.52992694773173	0.079321754853857\\
3.84296310818155	0.0832690793514853\\
4.11062060899258	0.0875780166168699\\
4.33964283651313	0.0921734841020688\\
4.53595960589226	0.0970026274988065\\
4.70468317609303	0.102025998783326\\
4.85016805807379	0.107212779799328\\
4.97609478747905	0.112538049196547\\
5.08555850940267	0.117981141872749\\
5.1811533942789	0.123524619191298\\
5.26504903195037	0.129153593038449\\
5.33905755369387	0.134855261019215\\
5.40469153146616	0.140618571027648\\
};
\addlegendentry{\# antennas M = 100}

\addplot [
mark=square,
mark repeat=5,
color=red,
solid] 
table[row sep=crcr]{%
0.565287620106576	0.0707604387169467\\
1.48693860859062	0.0538018177333006\\
2.44438341243013	0.0490921348057668\\
3.34610553701073	0.0478167822951984\\
4.16281228413	0.0480444435994544\\
4.88887225097113	0.0490910761581725\\
5.52837961959629	0.0506477520117273\\
6.08925424329429	0.0525515912482051\\
6.58055384939561	0.0547066414528412\\
7.01122533064798	0.0570513685035183\\
7.38954616299982	0.0595435755179557\\
7.72290909759617	0.0621527450257578\\
8.01777909967039	0.0648558651386862\\
8.27973462336714	0.0676350179653782\\
8.51354693751047	0.0704759137882256\\
8.7232729683445	0.0733669578290703\\
8.91234883570133	0.0762986292991627\\
9.08367768153062	0.0792630501920977\\
9.23970893653913	0.0822536732725987\\
};
\addlegendentry{\# antennas M = 200}

\addplot [
mark=*,
mark repeat=5,
color=darkgreen,
solid]
table[row sep=crcr]{%
0.711082843800288	0.0562522360773399\\
1.95086889309126	0.0410073687080199\\
3.29256340695234	0.0364457673758436\\
4.59326040449314	0.0348336444943308\\
5.79917804723123	0.0344876460717547\\
6.8938242102583	0.0348137684803262\\
7.87741506254118	0.0355446549124295\\
8.75752555146738	0.0365400018668951\\
9.54457175910796	0.0377177739437569\\
10.2495841733109	0.0390259734674474\\
10.883131060014	0.0404295416065158\\
11.4548376526097	0.0419036929685898\\
11.9732171447894	0.0434302655428161\\
12.445662718709	0.0449955950644699\\
12.8785188626251	0.0465892084641244\\
13.2771873100108	0.0482029804247359\\
13.6462433721066	0.0498305637278861\\
13.9895498926867	0.051466988253596\\
14.3103625235142	0.0531083680620389\\
};
\addlegendentry{\# antennas M = 400}

\end{axis}
\end{tikzpicture}
	\caption{Access latency (ms) as a function of the arrival rate for $K = 400$ and  for $M=100,200,400$.}
	\label{fig:Perf2}
\end{figure}

\tblue{Figure~\ref{fig:Perf1} and Figure~\ref{fig:Perf2} show the performance metrics for a scenario with K = 400 and M = 100, 200, 400 for an SNR of 10dB. The transmission slot duration is fixed to $\tau_u = 300$ and $\ tau_p = 100$: this ratio is chosen as it leads to a near-optimal solution (see above). We compute the average sum rate per device from which we determine the average delay to transmit 8 bytes per device over a bandwidth of 1MHz. This study relies on an information theory framework, where the devices are guaranteed to transmit their data reliably. Therefore, the average number of active devices that have successfully transmitted is also the average number of active users in the TTI. The performance metrics are plotted against the arrival rate per TTI.}

\section{Discussion and Comparison} 
\label{sec:discussion_and_comparison}

The shown performance results show that each solution provides a trade-off between throughput and latency. Yet we can conclude that very significant gains can be achieved if the following techniques are applied in the design of massive access protocols:
\begin{itemize}
	\item Physical layer: (i) \tblue{compressive} sensing for multi-user detection (CSMUD, CRAPLNC), (ii) multi-user decoding (OSTSAP, \tblue{CRAPLNC}), (iii) redesign of access preambles (OSTSAP) and (iv) multiple spatial layers (NOTAFT);
	\item Medium access layer: (v) coding over retransmissions (SBA, \tblue{CRAPLNC}), (vi) back-off schemes (OSTSAP, \tblue{CRAPLNC});
	\item Protocol Design: (vii) one-stage protocols (CSMUD) and (viii) low overhead network synchronization (NOTAFT).
\end{itemize}

One final remark is that for all the schemes a large part of the complexity is at the receiver of the base station, while the transmitter operation at the devices does not suffer an increase in complexity (with the exception of the CRPLNC scheme).


\section{Conclusions} 
\label{sec:conclusions}

For massive Machine Type Communications to take place, there is the need for efficient access protocols capable of withstanding a massive number of devices contending for network access. We have proposed several random-access schemes of one-stage and two-stages types. Several physical layer and medium access layer techniques have been considered. The physical layer techniques include multi-user detection using compressive sensing techniques, collision resolution and harness of interference using physical layer network coding and non-orthogonal access with relaxed time-alignment. The medium access layer techniques include coded random access and signature based access, one/two-stage random access and fast uplink access protocols with a focus on latency reduction. A common evaluation framework has been defined and individual performance results provided. These results will help on the design of a robust massive access solutions, by identifying which techniques lead to higher protocol performance, and doing so provide recommendations on the protocol design for the NR in 3GPP.


\section*{Acknowledgement}
This work has been performed in the framework of the Horizon 2020 Project FANTASTIC-5G under Grant ICT-671660, which is partly funded by the European Union. The authors would like to acknowledge the contributions of their colleagues in FANTASTIC-5G.

Emil Bj{\"o}rnson, Jesper H. S\o{}rensen and Erik G. Larsson are contributors of the work presented in section V.C.

\bibliographystyle{IEEEtran}
\bibliography{IEEEabrv,refs,refs_UB,refs_cttc,refs_HWDU,refs_AAU,refs_ccra,biblio,refsHHI}

\end{document}